\documentclass[final, 3p, times]{elsarticle}

\usepackage{amssymb}
\usepackage{amsmath}
\usepackage{tensor}
\usepackage{xcolor}
\usepackage{multirow}
\usepackage{float} 
\usepackage{tikz}
\usetikzlibrary{positioning}
\usetikzlibrary{calc}
\usepackage{hyperref}

\usepackage{etoolbox} 

\newcommand{\splitcelltab}[1]{\begin{tabular}{@{}c@{}}#1\end{tabular}} 

\newcommand{\wrt}{w.\,r.\,t.}
\newcommand{\eg}{e.\,g.}
\newcommand{\soth}{s.\,t.} 
\newcommand{\ie}{i.\,e.}

\newcommand{\oeda}{w.\,l.\,o.\,g.}

\newcommand{\resp}{resp.}
\newcommand{\cf}{cf.}

\newcommand{\formComma}{\,\text{,}}
\newcommand{\formPeriod}{\,\text{.}}

\newcommand{\surf}{\mathcal{S}}

\newcommand{\R}{\mathbb{R}}

\newcommand{\tangent}[2][]{\tensor{\operatorname{T}\!}{#1}#2}
\newcommand{\tangentS}[1][]{\tangent[#1]{\surf}}
\newcommand{\tangentR}[1][]{\tangent[#1]{\R^3\vert_{\surf}}}
\newcommand{\tangentQS}{\tensor{\operatorname{\mathcal{Q}}\!}{^2}\surf}
\newcommand{\tangentQR}{\tensor{\operatorname{\mathcal{Q}}\!}{^2}\R^3\vert_{\surf}}
\newcommand{\tangentCQR}{\tensor{\operatorname{C_{\surf}\mathcal{Q}}\!}{^2}\R^3\vert_{\surf}}

\newcommand{\paraC}{X}
\newcommand{\para}{\boldsymbol{\paraC}}
\newcommand{\normalC}{\nu}
\newcommand{\normal}{\boldsymbol{\normalC}}
\newcommand{\shopC}{I\!I}
\newcommand{\shop}{\boldsymbol{\shopC}}
\newcommand{\meanc}{\mathcal{H}}
\newcommand{\gaussc}{\mathcal{K}}

\newcommand{\dS}{\textup{d}\surf}

\newcommand{\eps}{\varepsilon}

\newcommand{\landau}{\mathcal{O}}

\newcommand{\Comp}{\mathsf{C}}

\newcommand{\nablaC}{\nabla_{\!\Comp}}

\newcommand{\rot}{\operatorname{rot}}
\newcommand{\Rot}{\operatorname{Rot}}

\newcommand{\Tr}{\operatorname{Tr}}
\renewcommand{\div}{\operatorname{div}}
\newcommand{\Div}{\operatorname{Div}}

\newcommand{\DivC}{\Div_{\!\Comp}}

\newcommand{\Grad}{\operatorname{Grad}}
\newcommand{\GradC}{\Grad_{\Comp}}

\newcommand{\DeltaC}{\Delta_{\Comp}}

\newcommand{\proj}{\operatorname{\Pi}}

\newcommand{\projQS}{\proj_{\tangentQS}}
\newcommand{\projQR}{\proj_{\tangentQR}}

\newcommand{\jau}{\mathcal{J}}
\newcommand{\timeD}[1][\Phi]{\operatorname{d}_{t}^{#1}\!}
\newcommand{\timeJ}{\mathfrak{J}}

\newcommand{\timeLQl}{\mathfrak{L}^{\mathcal{Q},\flat}}
\newcommand{\timeLu}{\mathfrak{L}^{\sharp}}
\newcommand{\timeLl}{\mathfrak{L}^{\flat}}

\newcommand{\Dt}[1][]{\operatorname{D}^{#1}_t \!}
\newcommand{\Dmat}{\Dt[\mfrak]}
\newcommand{\Dupp}{\Dt[\sharp]}
\newcommand{\Dlow}{\Dt[\flat]}
\newcommand{\Djau}{\Dt[\jau]}
\newcommand{\Dphi}{\Dt[\Phi]}

\newcommand{\potenergy}{\mathfrak{U}}
\newcommand{\energy}{\mathfrak{E}}
\newcommand{\fluxpotential}{\mathfrak{R}}

\newcommand{\energyK}{\energy_{\textup{K}}}

\newcommand{\Cset}{\mathcal{C}} 

\newcommand{\EL}{\textup{EL}}
\newcommand{\energyEL}{\potenergy_{\EL}} 
\newcommand{\HbEL}{\Hb_{\EL}}
\newcommand{\hbEL}{\hb_{\EL}}
\newcommand{\zetabEL}{\zetab_{\EL}}
\newcommand{\omegaEL}{\omega_{\EL}}
\newcommand{\SigmabEL}{\Sigmab_{\EL}}
\newcommand{\sigmabEL}{\sigmab_{\EL}}

\newcommand{\THT}{\textup{TH}} 
\newcommand{\energyTH}{\potenergy_{\THT}} 
\newcommand{\HbTH}{\Hb_{\THT}}
\newcommand{\hbTH}{\hb_{\THT}}
\newcommand{\omegaTH}{\omega_{\THT}}

\newcommand{\pTH}{p_{\THT}}

\newcommand{\BE}{\textup{BE}}
\newcommand{\energyBE}{\potenergy_{\BE}}

\newcommand{\fnorBE}{\fnor[\BE]}

\newcommand{\coeffIF}{\upsilon}

\newcommand{\IM}{\textup{IM}}
\newcommand{\energyIM}[1][\Phi]{\fluxpotential_{\IM}^{#1}}

\newcommand{\zetabIM}[1][\Phi]{\zetab_{\IM}^{#1}}

\newcommand{\SigmabIM}[1][\Phi]{\Sigmab_{\IM}^{#1}}
\newcommand{\sigmabIM}[1][\Phi]{\sigmab_{\IM}^{#1}}

\newcommand{\AC}{\textup{AC}}
\newcommand{\energyAC}{\fluxpotential_{\AC}} 
\newcommand{\HbAC}{\Hb_{\AC}}

\newcommand{\cAC}{\alpha}
\newcommand{\cIA}{\alpha_{\textup{I}}}
\newcommand{\cNA}{\alpha_{\textup{N}}}

\newcommand{\IA}{\textup{IA}}
\newcommand{\NA}{\textup{NA}}
\newcommand{\energyIA}{\fluxpotential_{\IA}} 
\newcommand{\energyNA}{\fluxpotential_{\NA}}
\newcommand{\FbIA}{\Fb_{\IA}}
\newcommand{\FbNA}{\Fb_{\NA}}

\newcommand{\NV}{\textup{NV}}
\newcommand{\energyNV}{\fluxpotential_{\NV}} 
\newcommand{\HbNV}{\Hb_{\NV}}
\newcommand{\hbNV}{\hb_{\NV}}

\newcommand{\SigmabNV}{\Sigmab_{\NV}}
\newcommand{\sigmabNV}{\sigmab_{\NV}}

\newcommand{\SC}{\textup{SC}}

\newcommand{\lambdabSC}{\lambdab_{\SC}} 

\newcommand{\zetabSC}{\zetab_{SC}}

\newcommand{\CB}{\textup{CB}}

\newcommand{\lambdaCB}{\lambda_{CB}}

\newcommand{\UN}{\textup{UN}}

\newcommand{\LambdabUN}{\Lambdab_{\UN}}
\newcommand{\lambdabUN}{\lambdab_{\UN}}
 
\newcommand{\lambdabotUN}{\lambda^{\bot}_{\UN}}

\newcommand{\NN}{\textup{NN}}

\newcommand{\lambdaNN}{\lambda_{\NN}}

\newcommand{\NF}{\textup{NF}}

\newcommand{\LambdabNF}{\Lambdab_{\NF}}

\newcommand{\IS}{\textup{IS}}

\newcommand{\LambdabIS}{\Lambdab_{\IS}}

\newcommand{\TOT}{\textup{Tot}}
\newcommand{\energyTOT}{\energy_{\TOT}}

\newcommand{\hil}{\operatorname{L}^{\!2}}
\newcommand{\hilspace}[1]{\hil(#1)}
\newcommand{\inner}[3][0]{\left\langle #3 \right\rangle_{\hspace{-#1pt}#2}}
\newcommand{\normsq}[3][0]{\left\| #3 \right\|_{\hspace{-#1pt}#2}^2}
\newcommand{\innerH}[3][0]{\inner[#1]{\hilspace{#2}}{#3}}
\newcommand{\normHsq}[3][0]{\normsq[#1]{\hilspace{#2}}{#3}}
\newcommand{\normHQRsq}[2][0]{\normHsq[#1]{\tangentQR}{#2}}

\newcommand{\fa}{\textup{fa}}
\newcommand{\sigmabfa}{\sigmab_{\fa}}
\newcommand{\hbfa}{\hb_{\fa}}

\newcommand{\ceven}{\alpha_{\textup{e}}}
\newcommand{\codd}{\alpha_{\textup{o}}}

\newcommand{\Wb}{\boldsymbol{W}}
\newcommand{\wb}{\boldsymbol{w}}
\newcommand{\wnor}{w_{\bot}}
\newcommand{\qb}{\boldsymbol{q}}
\newcommand{\Qb}{\boldsymbol{Q}}
\newcommand{\rb}{\boldsymbol{r}}
\newcommand{\Rb}{\boldsymbol{R}}

\newcommand{\Gb}{\boldsymbol{G}}
\newcommand{\Sb}{\boldsymbol{S}}
\newcommand{\Ab}{\boldsymbol{A}}
\newcommand{\Eb}{\boldsymbol{E}}

\newcommand{\Vb}{\boldsymbol{V}}
\newcommand{\vb}{\boldsymbol{v}}
\newcommand{\vnor}{v_{\bot}}
\newcommand{\lambdab}{\boldsymbol{\lambda}}
\newcommand{\Lambdab}{\boldsymbol{\Lambda}}

\newcommand{\sigmab}{\boldsymbol{\sigma}}
\newcommand{\etab}{\boldsymbol{\eta}}
\newcommand{\nullb}{\boldsymbol{0}}
\newcommand{\pb}{\boldsymbol{p}}

\newcommand{\Sigmab}{\boldsymbol{\Sigma}}
\newcommand{\Hb}{\boldsymbol{H}}

\newcommand{\Fb}{\boldsymbol{F}}
\newcommand{\fb}{\boldsymbol{f}}
\newcommand{\fnor}[1][]{f^{\bot}_{#1}}

\newcommand{\hb}{\boldsymbol{h}}
\newcommand{\zetab}{\boldsymbol{\zeta}}

\newcommand{\ab}{\boldsymbol{a}}
\newcommand{\anor}{a_{\bot}}

\newcommand{\Cb}{\boldsymbol{C}}

\newcommand{\kb}{\boldsymbol{k}}

\newcommand{\txi}{\tilde{\xi}}

\newcommand{\Gbcal}{\boldsymbol{\mathcal{G}}}
\newcommand{\Abcal}{\boldsymbol{\mathcal{A}}}

\newcommand{\Vcal}{\mathcal{V}} 

\newcommand{\Id}{\boldsymbol{Id}}
\newcommand{\IdS}{\Id_{\surf}}

\newcommand{\Ib}{\boldsymbol{I}}
\newcommand{\IbxiQ}{\Ib_{\xi}[\Qb]}

\newcommand{\IbtxiQ}{\Ib_{\tilde{\xi}}[\Qb]}

\newcommand{\shopQ}{\shop_{\mathcal{Q}}}
\newcommand{\hodge}{*\!}
\newcommand{\halfhodge}{*^{\!\frac{1}{2}}}

\newcommand{\mfrak}{\mathfrak{m}}
\newcommand{\ofrak}{\mathfrak{o}}

\newcommand{\deltafrac}[2]{\frac{\delta #1}{\delta #2}}
\newcommand{\ddfrac}[2][]{\frac{\textup{d} #1}{\textup{d} #2}}
\newcommand{\ddt}[1][]{\ddfrac[#1]{t}}

\newcommand{\dbdot}{\operatorname{:}}

\makeatletter
\patchcmd{\math@cr@@@align}{\cr}{\global\let\df@label\@empty\cr}{}{}
\makeatother

\begin{document}
	
\begin{frontmatter}

\title{Active nematodynamics on deformable surfaces}

\author[1]{Ingo Nitschke\corref{cor1}}
\cortext[cor1]{Corresponding author: ingo.nitschke@tu-dresden.de}
\author[1,2,3]{Axel Voigt}

\address[1]{Institut f{\"u}r Wissenschaftliches Rechnen, Technische Universit{\"a}t Dresden, 01062 Dresden, Germany}
\address[2]{Dresden Center for Computational Materials Science (DCMS), Technische Universit{\"a}t Dresden, 01062 Dresden, Germany}
\address[3]{Center for Systems Biology Dresden (CSBD), Pfotenhauerstr. 108, 01307 Dresden, Germany}

\journal{arXiv}

\begin{abstract}
We consider active nematodynamics on deformable surfaces. Based on a thermodynamically consistent surface Beris-Edwards model we add nematic activity and focus on the emerging additional coupling mechanism between the nematic field, the flow field and the curved surface. We analyse the impact of the active nematic force at topological defects. Under the presence of curvature all defects become active and contribute not only tangential forces but also normal forces. This confirms the proposed role of topological defects in surface evolution and provides the basis for a dynamic description of morphogenetic processes.
\end{abstract}

\end{frontmatter}


\section{Introduction}

Recent theoretical and experimental studies demonstrate that nematic-like ordering plays a fundamental role in morphogenesis \cite{Maroudas-Sacksetal_NP_2021,Vafaetal_PRL_2022,Wangetal_PNAS_2023}. 
The microscopic constituents responsible for this ordering are elongated cells forming the tissue \cite{saw2017topological}, or actin–myosin fibres underneath the tissue \cite{Maroudas-Sacksetal_NP_2021}. These constituents tend to locally align parallel to each other and form long-ranged orientational order \cite{de_Gennes_book}. On curved surfaces the cells or fibres not only align with respect to each other but also respond to the curvature and align with the principle curvature directions of the surface \cite{napoli2012surface,golovaty2017dimension,Nitschke_2018}. If the surface is deformable this interaction between geometric properties of the surface and the orientational order can lead to shape changes \cite{Park_EPL_1992,Nitschke_2020}. For active systems, where the constituents consume energy and generate forces, additional coupling mechanisms between geometric properties and orientational order emerge, giving rise to interesting dynamic
behaviours, including buckling and wrinkling instabilities \cite{Senoussi_2019,Str_bing_2020}. If topological defects are present in the nematic field, the coupling becomes even more involved. Similar to out-of-plane deformation of thin crystalline sheets near dislocations \cite{Lehtinen2013,Zhang2014a,benoitmaréchal2024mesoscale}, also topological defects in tissues lead to localized shape changes \cite{kawaguchi2017topological,saw2017topological,guillamat2022integer}. While these mechanisms in principle provide enough physics to describe the role of nematic-like ordering in the morphology of protrusions and extrusions, or in morphogenesis in general, all previous attempts to link theoretical and experimental studies are limited to selected snapshots of the evolution \cite{Metselaar_2019,Maroudas-Sacksetal_NP_2021,hoffmann2022theory,Wangetal_PNAS_2023}. 

A quantitative dynamic description of morphogenesis using these mechanism is still in its infancy. Experimental studies to unveil the physics of the dynamics and corresponding dynamic modeling is restricted to special situations. Nematics-like ordering on spherical vesicles is one example. Here the geometry is fixed and its topology requires a net topological charge of $+2$, which in equilibrium is realized by four $+1/2$ defects arranged in a tetrahedral configuration \cite{lubensky1992orientational,nelson2002toward}. For active systems the four defects are motile and oscillate between this and a planar configuration, which drives spontaneous flows \cite{Keberetal_Science_2014,Alaimoetal_SR_2017,henkes2018dynamical}. Also nematic fields on non-spherical geometric stationary surfaces have been investigated with respect to dynamics of the topological defects \cite{ellis2018curvature,Nestler_2022}. Similar results also exist for epithelial acini \cite{tanner2012coherent,wang2013rotational}, which are spherical-like structures with the epithelium surrounding a lumen. In these systems not only cell shapes adjust to curvature, they also lead to coordinated rotational movement \cite{happel2022effects,brandstatter2021curvature}. Persistent rotational movement has also been investigated on cylindrical surfaces using experimental \cite{glentis2022emergence} and computational \cite{happel2024coordinated} studies. Using the elongation of the cells to define surface nematic fields and considering nematic defects in such systems shows defect binding and unbinding \cite{happel2022effects}. Other simplified cases neglect either the hydrodynamic interactions or the nematic field, but allow for variations of the shape. Different models allowing for the simultaneous relaxation of the surface nematic field and the surface itself have been proposed \cite{Nitschke_2020,NitschkeSadikVoigt_A_2022}. The resulting equilibrium shapes and defect configurations strongly depend on model assumptions. In addition also the dynamics varies according to constitutive descriptions for the immobility mechanism utilizing different time derivatives \cite{stone2023note,NitschkeSadikVoigt_A_2022,NitschkeVoigt_2023}. Neglecting the nematic field but keeping the hydrodynamics lead to models for fluid deformable surfaces \cite{Torres-Sanchez_2019,Reuther_2020,Krause_2023}, which consider surface (Navier-)Stokes equations with bending forces. Numerical simulations demonstrate an enhanced evolution compared to classical Helfrich-type models \cite{Helfrich}. However, activity is typically not considered in these approaches.

All quantitative comparisons between experimental and theoretical studies for such special situations demonstrate the importance of nematic and hydrodynamic interactions \cite{reymann2016cortical,Nestler_2022,bhatnagar2023axis}. However, combining these approaches to consider all relevant mechanical phenomena in one consistent dynamic model for morphogenesis is a challenging task. This paper addresses this issue. Some attempts in this direction already exist, they are summarized in \cite{al2021active}. In \cite{Metselaar_2019} the possibility to form protrusions and extrusions as a result of active nematodynamics is demonstrated. But instead of a surface model a three-dimensional model is considered and non of the details mentioned above are discussed, see Section \ref{sec:existing} for further discussions. A very detailed investigation of the mechanics of active surfaces and nematic broken symmetries can be found in \cite{salbreux2017mechanics,Salbreux_2022}. However, a dynamic model which combines these forces is not formulated. Closest to our goal comes the approach in \cite{Al_Izzi_2023}. 
They develop a hydrodynamic theory of active nematic fluids on deformable surfaces. However, they focus on the Frank-Oseen free energy, which is based on a director field, and only consider the effect of activity on the surface morphology around a $+1$ topological defect. We follow a different perspective and consider a Q-tensor (Landau-de Gennes) theory, which also allows for $+1/2$ and $-1/2$ defects. Our investigation is based on the recently proposed surface Beris-Edwards models \cite{Nitschke2023}. These models, considering nematodynamics on deformable surfaces, provide a thermodynamically consistent model taking the most relevant interactions between the nematic field, the flow field and the curved surface into account and address implications on model assumptions, such as surface conformity, uniaxiality and immobility mechanisms. The models are derived using the Lagrange-D'Alembert approach. We here only add nematic activity and focus on the emerging additional coupling mechanisms.

The paper is structured as follows: In Section \ref{sec:passivBE} we introduce notation and review the (passive) surface Beris-Edwards models. We consider a general description with possible constraints introduced by Lagrange multipliers. In Section \ref{sec:activeBE} we consider nematic activity. Based on an active flux potential the model is extended by active forces leading to active surface Beris-Edwards models. We formulate, as in the passive case, a general model, but also consider a surface conforming model. The last allows for direct comparison with existing special cases. In Section \ref{sec:diss} we analyse the impact of the active nematic force at topological defects, essentially demonstrating that under the presence of curvature all relevant defects, $+ 1/2$ defects but also $- 1/2$, $+ 1$ and $- 1$ defects become active and contribute not only tangential forces but also normal forces. We also draw conclusions in this section. Technical details and a comparison of different flow alignment mechanisms are considered in \ref{app:a} and \ref{sec:flow_alignment}, respectively. 

\section{Passive Surface Beris-Edwards models} \label{sec:passivBE}

\subsection{Notation and Mathematical Preliminaries}

Since we strictly adhere to the notation and preliminaries in \cite{Nitschke2023} we here provide only an essential introduction necessary for this paper.
We assume a sufficiently smooth parameterizable moving surface $ \surf\subset\R^3 $ in space and time.
Building on this, we consider time-dependent Euclidean-based $ n $-tensor fields in $ \tangentR[^n] $.
We call $\tangentR[^0] = \tangentS[^0]$ the space of scalar fields, $ \tangentR[^1]=\tangentR $ the space of vector fields, and $ \tangentR[^2] $ the space of 2-tensor fields.
Important subtensor fields are tangential n-tensor fields in $ \tangentS[^n] \le \tangentR[^n] $ and (biaxial) Q-tensor fields in $ \tangentQR < \tangentR[^2] $.
The latter space in turn comprises surface-conforming Q-tensor fields in $ \tangentCQR < \tangentQR $ and tangential (flat-degenerated)   Q-tensor fields in $ \tangentQS < \tangentCQR $.
More constructive: 
$ \tangentS[^1] = \tangentS = \{ \Rb\in\tangentR \mid \Rb\normal = \nullb \} $;
$ \tangentS[^2] = \{ \Rb\in\tangentR[^2] \mid \Rb\normal = \normal\Rb = \nullb \} $;
$ \tangentQR = \{ \Rb\in\tangentR[^2] \mid \Rb^T = \Rb \text{ and }  \Tr\Rb = 0 \} $;
$ \tangentCQR = \{ \Qb \in \tangentQR \mid \exists\lambda\in\tangentS[^0] : \Qb\normal = \lambda\normal \} $;
$ \tangentQS = \{ \rb\in\tangentS[^2] \mid \rb^T = \rb \text{ and }  \Tr\rb = 0 \} = \{ \Qb\in\tangentCQR \mid \Qb\normal=\nullb \} $,
where $\normal\in\tangentR$ is the surface normal field.
On tangential tensor fields we use the covariant derivative $ \nabla: \tangentS[^n] \rightarrow \tangentS[^{n+1}] $ and its common derived differential operators, 
like the covariant divergence $ \div=\Tr\circ\nabla=-\nabla^*: \tangentS[^n] \rightarrow \tangentS[^{n-1}] $.
On more general $ n $-tensor fields we use the componentwise surface derivative $ \nablaC: \tangentR[^n] \rightarrow  \tangentR[^n]\otimes\tangentS $,
which is basically the scalar-valued covariant derivative on its Cartesian proxy component fields, see \cite{NitschkeSadikVoigt_A_2022,NitschkeVoigt_2023,BachiniKrauseNitschkeVoigt_2023, Nitschke2023}.
The componentwise trace-divergence is $ \DivC = \Tr\circ\nablaC: \tangentR[^n] \rightarrow \tangentR[^{n-1}] $.
Note that only on right-sided tangential $ n $-tensor fields $ \tangentR[^{n-1}]\otimes\tangentS $ holds the $ L^2 $-adjoint relation $ \DivC=-\nablaC^* $.
In our models this is always the case for stress tensor fields, \ie\ it holds $ \DivC(\sigmab + \normal\otimes\zetab) = -\nablaC^*(\sigmab + \normal\otimes\zetab) $
for $ \sigmab\in\tangentS[^2] $ and $ \zetab\in\tangentS $, but it is not valid for more general 2-tensor fields in $ \tangentR[^2] $.
On the other hand, we could exploit this circumstance and define the adjoint componentwise gradient $ \GradC:=-\DivC^* $.
We use this operator solely for scalar fields $ f\in\tangentS[^0] $, where $\GradC f = \DivC (f\IdS) = \nabla f + \meanc f \normal$ holds, with mean curvature $ \meanc = \Tr\shop $,
(tensor-valued) second fundamental form\footnote{Other names are (extended) Weingarten map or shape operator.} $ \shop = -\nablaC\normal \in\tangentS[^2] $,
and  surface identity tensor $ \IdS\in\tangentS[^2] $,  \ie\ $ \IdS\Wb $ is the tangential part of the vector field $ \Wb\in\tangentR $.
Based on the derivative $ \nablaC $ on vector fields $ \Wb=\wb + \wnor\normal\in\tangentS\oplus(\tangentS[^0])\normal=\tangentR $, we introduce a few recurring quantities, which are
the surface 
deformation\footnote{The naming relates to small surface deformations ``$ \surf + \eps\Wb $'', see \cite{NitschkeSadikVoigt_A_2022}.}
gradient and its tangential part
\begin{align}
    \Gbcal[\Wb] \label{eq:Gbcal}
        &:= \nablaC\Wb - \normal\nablaC\Wb\otimes\normal
        = \Gb[\Wb] + \normal\otimes\left( \nabla\wnor + \shop\wb \right) - \left( \nabla\wnor + \shop\wb \right)\otimes\normal \formComma  \\
    \Gb[\Wb] \label{eq:Gb}
        &:= \IdS\nablaC\Wb
          = \nabla\wb - \wnor\shop \formPeriod
\end{align} 
Their symmetric and skew-symmetric parts are
\begin{align}
    \Sb[\Wb] \label{eq:Sb}
        &:= \frac{1}{2}\left( \Gbcal[\Wb] + \Gbcal^T[\Wb] \right) 
          = \frac{1}{2}\left( \Gb[\Wb] + \Gb^T[\Wb] \right)  
          = \frac{1}{2}\left( \nabla\wb + (\nabla\wb)^T \right) - \wnor\shop \formComma \\
    \Abcal[\Wb]  \label{eq:Abcal}
      &:= \frac{1}{2}\left(\Gbcal[\Wb] - \Gbcal^{T}[\Wb]\right) 
        =  \Ab[\Wb] + \normal\otimes\left( \nabla\wnor + \shop\wb \right) - \left( \nabla\wnor + \shop\wb \right)\otimes\normal \formComma \\
    \Ab[\Wb] \label{eq:Ab}
          &:= \frac{1}{2}\left(\Gb[\Wb] - \Gb^{T}[\Wb]\right)
          = \frac{1}{2}\left( \nabla\wb - (\nabla\wb)^T \right)
          = -\frac{\rot\wb}{2}\Eb \formComma
\end{align}
where $ \Eb\in\tangentS[^2] $ is the Levi-Civita tensor, \ie\ $ -\Eb\wb= *\wb $ gives the tangential Hodge-dual of $ \wb $,
and $\rot\wb = - \Eb\dbdot\nabla\wb $ the curl of $ \wb $.

The kinematic of $ \surf $ can be characterized by the observer velocity $ \Vb_{\!\ofrak}\in\tangentR $ \wrt\ any valid surface observer, see \cite{NitschkeVoigt_JoGaP_2022}.
Within a spatial discretization this observer velocity could serve as the grid velocity for instance.
However, from a physical point of view we are only interested in the material velocity $ \Vb\in\tangentR $, which determines the motion of the material.
The only mandatory relation between observer and material velocity is $ \Vb_{\!\ofrak}\normal = \Vb\normal =: \vnor $, \ie\ the tangential part $ \vb_{\!\ofrak} = \IdS\Vb_{\!\ofrak} $ of the observer velocity is still arbitrary.
Conventional choices are  $\Vb_{\!\ofrak}=\Vb$ (material/Lagrangian perspective) and $ \vb_{\!\ofrak} = \nullb $ (transversal/tangential-Eulerian perspective).
In this paper, the observer velocity is only important to determine local observer-invariant tensor rates sufficiently, 
\eg\ $ \Dmat\Vb = \partial_t\Vb + (\nablaC\Vb)(\Vb-\Vb_{\!\ofrak}) $ is the material acceleration and 
$ \Dmat\Rb = \partial_t\Rb + (\nablaC\Rb)(\Vb-\Vb_{\!\ofrak}) $ the material tensor rate of $ 2 $-tensor fields $ \Rb\in\tangentR[^2] $. 
Other tensor rates can be derived from the material rate, \eg\ 
$ \Djau\Rb = \Dmat\Rb - \Abcal[\Vb]\Rb + \Rb\Abcal[\Vb] $
is the Jaumann/corotational rate and 
\begin{align}
   \Dlow\Rb 
        = \Dmat\Rb + \Gbcal^T[\Vb]\Rb + \Rb\Gbcal[\Vb]
        = \Djau\Rb + \Sb[\Vb]\Rb + \Rb\Sb[\Vb]  \label{eq:Dlow_rel_to_Djau}
\end{align}
the lower-convected rate of 2-tensor fields $ \Rb\in\tangentR[^2] $.
Moreover, tensor rates can be orthogonally decompose \wrt\ tangential and normal spaces, \eg $\,$
\begin{align}\label{eq:Dupp}
 \forall \, \Rb 
    =\rb+r_{\bot}\normal\in\tangentR:  
 \quad \Dupp\Rb 
    = \timeLu\rb + \dot{r}_{\bot}\normal
\end{align}
is the upper-convected vector field rate, where $ \timeLu:\tangentS \rightarrow \tangentS $ is the tangential upper-convected time derivative on tangential vector fields.
See \cite{NitschkeVoigt_2023} for more details, more observer-invariant \mbox{(sub-)}tensor rates, relations between them and their orthogonal decompositions.

\subsection{Passive Energetic Contributions}\label{sec:energetic_contributions}

The fundament of the derivation of the surface Beris-Edwards model in \cite{Nitschke2023} is the Lagrange-d'Alembert principle.
It is based on thermodynamical consistent  considerations of energies and energy flux potentials.
This includes the kinetic energy for point masses
\begin{align*}
\energyK 
    &:= \frac{1}{2} \innerH{\tangentR}{\rho \Vb, \Vb} \formComma
\end{align*}
where $ \rho\in\tangentS[^0] $ is the material mass density and $ \Vb\in\tangentR $ the material velocity field;
potential energies
\begin{align*}
    \energyEL
             &:= \frac{L}{2}\normHsq{\tangentQR\otimes\tangentS}{\nablaC\Qb} \formComma\\
    \energyTH
            &:= \innerH{\tangentR[^2]}{a\Qb + \frac{2b}{3}\Qb^2 + c\Qb^3, \Qb}\formComma\\
    \energyBE
            &:= \frac{\kappa}{2} \normHsq{\tangentS[^0]}{\meanc - \meanc_{0}}\formComma
\end{align*}
where $ \energyEL $ is the one-constant elastic, $ \energyTH $ the thermotropic and $ \energyBE $ the surface bending energy with $ \Qb\in\tangentQR $ the Q-tensor field and $\meanc$ the mean curvature and parameters: one-constant elastic parameter $L \geq 0$, thermotropic coefficients $a, b,$ and $c$, bending stiffness $\kappa \geq 0$ and spontaneous curvature $\meanc_0 \in \R$; and dissipation potentials
\begin{align*}
    \energyIM
            &:= \frac{M}{2}\normHQRsq{\Dphi\Qb} \formComma\\
    \energyNV 
            &:= \frac{\coeffIF}{4}\normHsq{\tangentR[^2]}{\Dlow\left(\Id - \xi\Qb\right)} \formComma
\end{align*}
where $ \energyIM $ is the immobility and $  \energyNV $ the nematic viscous flux potential with immobility coefficient $M \geq 0$ and isotropic viscosity $v \geq 0$.
The symbol $\Phi$ determines whether we consider a material ($\Phi=\mfrak$) or a corotational/Jaumann ($\Phi=\jau$) immobility mechanism.

\subsection{Passive Dynamic Equations}

In \cite{Nitschke2023} these energies and energy flux potentials are considered in detail resulting in a general (passive) surface Beris-Edwards model, which reads:

\textit{Find the material velocity field $ \Vb\in\tangentR $, Q-tensor field $ \Qb\in\tangentQR $,  surface pressure field $ p\in\tangentS $ and Lagrange parameter fields $\Lambdab_{\gamma}\in\Vcal_{\gamma}\le\tangentR[^{n_{\gamma}}]$ for all $ \gamma\in\Cset$ \soth}
\begin{subequations}\label{eq:model_lagrange_multiplier}
\begin{gather}
    \rho\Dmat\Vb \label{eq:model_lagrange_multiplier_floweq}
			=  \GradC\left(\pTH - p\right) 
				+ \fnorBE \normal
                + \DivC \widetilde{\Sigmab}
				+ \sum_{\gamma\in\Cset} \Fb_{\gamma} \formComma \\
    \widetilde{M} \Dt[\Phi]\Qb \label{eq:model_lagrange_multiplier_moleculareq}
                 = \HbEL + \HbTH + \widetilde{\Hb}^{\Phi}_{\NV} + \sum_{\gamma\in\Cset} \Hb_{\gamma} \formComma\\
    0 = \DivC\Vb \formComma\quad\quad
    \nullb = \Cb_{\gamma}, \quad \forall\gamma\in\Cset \label{eq:model_lagrange_multiplier_constraineq}
\end{gather}
\end{subequations}
\textit{holds for $ \dot{\rho}=0 $ and given initial conditions for $ \Vb $, $ \Qb $ and mass density $ \rho\in\tangentS[^0] $.}

The passive stress field is summarized to
\begin{align*}
   \widetilde{\Sigmab} &= \SigmabEL + \SigmabIM + \SigmabNV^0 + \xi\SigmabNV^1 + \xi^2\SigmabNV^2\formComma
\end{align*}
and the nematic viscosity Q-tensor force partly to
\begin{align*}
\widetilde{\Hb}^{\Phi}_{\NV} = \xi(\HbNV^1 + \xi\widetilde{\Hb}^{2,\Phi}_{\NV})\formComma
\end{align*}
where $ \xi\in\R $ is the anisotropy coefficient. To determine all terms sufficiently, mandatory quantities are given in Table \ref{tab:forces} and optional constraint quantities, such as $ \Fb_{\gamma}, \Hb_{\gamma}, \Cb_{\gamma}$ and $\Vcal_{\gamma}$, are given in Table \ref{tab:constraints} within \ref{app:a}.
The considered nematic anisotropy results in flow alignment with the Q-tensor field. For a brief comparison of different approaches to achieve this, see Appendix \ref{sec:flow_alignment}. The immobility coefficient adapted to the nematic viscosity is $ \widetilde{M}=  M + \frac{\coeffIF\xi^2}{2}  $.
The symbol $ \Phi $ determine the preferred immobility mechanism.
We consider either the Jaumann model ($ \Phi=\jau $) or the material model ($\Phi=\mfrak$).
Restrictions can be chosen from the set $ \Cset \subset \{ \SC, \CB, \UN, \IS, \NN, \NF \} $ arbitrarily except for combinations that are mutually exclusive or merge into one another. These individual constraints relate to surface conformity ($ \SC $), a constant eigenvalue $ \beta $ in normal direction ($ \CB $), uniaxility ($ \UN $), an isotropic state ($ \IS $), no flow in normal direction ($ \NN $) and no flow at all ($ \NF $). 
Inextensibility is mandatory and already considered.
Index abbreviations refer to elasticity ($ \EL $), thermotropism ($ \THT $), bending ($\BE$),
immobility ($ \IM $), and nematic viscosity ($ \NV $).

Eqs. \eqref{eq:model_lagrange_multiplier} provide a thermodynamically consistent formulations. As demonstrated in \cite{Nitschke2023}, the total energy rate yields
\begin{align*}
    \ddt\left( \energyK + \energyEL + \energyTH + \energyBE \right)
        &= -2\left(  \energyIM + \energyNV \right)
        \le 0 \formComma
\end{align*}
regardless of any additional constraints on the Q-tensor field or material velocity field. 

The dynamics allows for the simultaneous relaxation of the surface Q-tensor field and the shape of the surface by taking the tight interplay between geometry and flow field and flow field and Q-tensor field into account. Already the most extreme restrictions show these couplings. An isotropic fluid, $\IS \in \Cset$ and no further constraints, leads to the model for a fluid deformable surface \cite{Arroyo_2009,Torres-Sanchez_2019,Reuther_2020,Krause_2023}. This model combines the surface Navier-Stokes equations in tangential and normal direction \cite{Reuther_2015,Koba_2017,Reuther_MMS_2018,Koba_2018,Miura_2018,Jankuhn_2018} with bending forces resulting from Helfrich/Willmore energies \cite{Helfrich}. As a result of the tight interplay between geometry and flow field, in the presence of curvature, any shape change is accompanied by a tangential flow and, vice-versa, the surface deforms due to tangential flow. The other extreme case is the situation without flow. The overdamped limit of Eqs. \eqref{eq:model_lagrange_multiplier}, with additional simplifications, leads to a surface Landau-de Gennes-Helfrich model, which allows for simultaneous relaxation of the Q-tensor field and the surface. Depending on the constraints on the Q-tensor field different equilibrium states emerge. While purely intrinsic models, e.g. $\CB \in \Cset$ with $\beta = 0$ lead to a tetrahedral defect arrangement \cite{Park_EPL_1992}, taking extrinsic curvature contributions into account, e.g. $\CB \in \Cset$ with $\beta > 0$ lead to asymmetric shapes with a planar defect arrangement \cite{Nitschke_2020}. This difference can be attributed to the tendency of the director field to align with principle curvature directions if $\beta > 0$. These coupling terms are only two of many present in the (passive) surface Beris-Edwards models \cite{Nitschke2023}.

\section{Nematic Activity} \label{sec:activeBE}

In this section we take a step further and additionally consider an unbounded flux potential. The resulting active forces provide additional coupling terms between the surface Q-tensor field, the shape of the surface and the flow field leading to various additional implications.

\subsection{Active Flux Potential}\label{sec:active_flux_potential}

We consider the nematic metric
\begin{align}\label{eq:nematic_metric}
	\IbxiQ &:= \Id - \xi\Qb \in\tangentR[^2] ,
\end{align}
where $ \xi\in\R $ is the anisotropy coefficient.
This states a simple linear expansion for an anisotropic metric depending on Q-tensor fields $ \Qb\in\tangentQR $.
Furthermore, its temporal distortion is fully quantified by the lower-convected rate $ \Dlow\IbxiQ\in\tangentR[^2] $, 
since the scalar rate $ \ddt\IbxiQ(\Rb_1,\Rb_2) = (\Dlow\IbxiQ)(\Rb_1,\Rb_2) $ of the associated anisotropic local inner product 
is invariant with respect to the choice of vector fields $ \Rb_1,\Rb_2\in\ker\Dupp < \tangentR $ frozen in the material flow.
We recapitulate from \cite{NitschkeVoigt_2023} and Eq. \eqref{eq:Dupp} that 
\begin{align*}
	\Rb&\in\ker\Dupp 
		&&\Longleftrightarrow &\Dupp\Rb &=\nullb
		&&\Longleftrightarrow &\timeLu\rb &=\nullb \quad\text{and}\quad \dot{r}_{\bot}=0
\end{align*}
holds for all vector fields $ \Rb = \rb + r_{\bot}\normal \in \tangentR $.
The second moment of Eq. \eqref{eq:nematic_metric}, $ \Tr(\Dlow\IbxiQ)^2 = \|\Dlow\IbxiQ\|^2 $ leads to nematic viscosity, see \cite{Nitschke2023}. We now investigate the first moment $\Tr(\Dlow\IbxiQ)$ in the same manner.
For this purpose we define the flux potential
\begin{align}
	\energyAC 
		&:= \frac{\cAC}{2}\int_{\surf} \Tr(\Dlow\IbtxiQ)\dS \label{eq:fluxAC_origin}
\end{align}
where $ \cAC\in\R $.
We rename the anisotropy coefficient $ \xi $ to $ \txi $ specifically only for this potential to further separate the anisotropic viscous influence from the active one.
This allows us to compare the resulting model also with models considering only anisotropic active contributions by neglecting anisotropic viscosity.
However, we recommend to set $ \xi=\txi $ if both contributions are treated, since in our opinion $ \xi $ is a material quantity and should be used equally in both cases.
Since the Jaumann derivative is $ \R^3 $-metric compatible, \ie\ $ \Djau\Id=\nullb $, it holds 
$ \Dlow\IbtxiQ = -\txi\Djau\Qb + \Sb[\Vb]\IbtxiQ + \IbtxiQ\Sb[\Vb] $ by Eq. \eqref{eq:Dlow_rel_to_Djau}. 
The space of Q-tensor fields $ \tangentQR $ is closed by the Jaumann derivative \cite{NitschkeVoigt_2023}, \ie\ $ \Tr(\Djau\Qb) = 0 $ particularly.
Therefore the flux potential \eqref{eq:fluxAC_origin} results in
\begin{align}
	\energyAC 
		&= \cAC \innerH{\tangentR[^2]}{\IbtxiQ,\Sb[\Vb]}
		= \cAC \innerH{\tangentR[^2]}{\IdS - \txi \IdS\Qb,\nablaC\Vb} \formPeriod
	\label{eq:fluxAC}
\end{align}
Contrarily to the passive flux potentials in \cite{Nitschke2023}, this potential is neither quadratic, nor generally positive at all.
Moreover it transfers one-to-one into energy exchange with a sink/source, \eg\ by the surrounded environment or internal energies, see Section \ref{sec:energy_rate}.
Therefore, Eq. \eqref{eq:fluxAC} does not stipulate a passive mechanism, hence we call it the active flux potential.
However, as for the nematic viscosity, it is rigid body motion invariant, since $ \Sb[\Vb] $ is linear in $ \Vb $ and it holds $ 2\Sb[\Vb] = \frac{1}{2}\Dlow\IdS = \nullb $ if and only if $ \Vb $ describes a rigid body motion.

For a better classification, we decompose Eq. \eqref{eq:fluxAC} into an isotropic and a nematic part
\begin{align}
	\energyIA 
		&:= \cIA \int_{\surf} \DivC\Vb\dS
		&&= \cIA \innerH{\tangentR[^2]}{\IdS,\nablaC\Vb} \formComma \label{eq:fluxIA}\\
	\energyNA
		&:= \cNA \innerH{\tangentQR}{\Qb , \projQR\Sb[\Vb]}
		&&= \cNA\innerH{\tangentR[^2]}{\IdS\Qb,\nablaC\Vb} \label{eq:fluxNA}
\end{align}
with $ \cIA=\cAC $, $ \cNA = -\txi\cAC $, and $\energyAC = \energyIA + \energyNA$.
The active isotropic flux potential $ \energyIA $ solely depends on the surface extensibility $ \DivC\Vb $, 
and the active nematic flux potential $ \energyNA $ on the deviatoric stress $ \projQR\Sb[\Vb] = \Sb[\Vb] - \frac{\DivC\Vb}{3}\Id $ with respect to $ \Qb $.
Hence the active isotropic flux potential vanishes for inextensible fluids.
However, it cannot be neglected if one ensure $ \DivC\Vb=0 $ by the Lagrange-multiplier method, as done in \cite{Nitschke2023}.
Eventually, the isotropic activity leads to an additional pressure-like force as we see below, which does not have any impact on divergence-free solutions $ \Vb $.


\subsection{Active Forces}\label{sec:active_forces}

Since the active flux potential is linear \wrt\ the material velocity $ \Vb $, the active isotropic and nematic fluid force fields $ \FbIA,\FbNA\in\tangentR $ are easy to compute.
The variations 
\begin{align*}
\innerH{\tangentR}{\deltafrac{\energyIA}{\Vb},\Wb} 
	&=: -\innerH{\tangentR}{\FbIA,\Wb}  \\
\text{and}\quad\innerH{\tangentR}{\deltafrac{\energyNA}{\Vb},\Wb} 
	&=: -\innerH{\tangentR}{\FbNA,\Wb}   
\end{align*} 
of Eqs. \eqref{eq:fluxIA} and \eqref{eq:fluxNA} in arbitrary displacement direction fields $ \Wb\in\tangentR $ yield
\begin{align}
	\FbIA \label{eq:active_geometric_force}
		&= \cIA\DivC\IdS
		 = \cIA\GradC 1
		 = \cIA\meanc\normal \formComma \\
	\FbNA \label{eq:active_nematic_force}
		&= \cNA\DivC\left( \IdS\Qb\IdS \right) \formComma
\end{align}
respectively. The active flux potential does not depend on any Q-tensor rates, which could serve as a process variable associated to $ \Qb $.
As a consequence no active forces $ \HbAC\in\tangentQR $ occur in the Q-tensor equation, \resp\ we stipulate $ \Hb_{\IA} = \Hb_{\NA} = \nullb $ in compliance with the setup in \cite{Nitschke2023}. In other words these active forces only enter Eq. \eqref{eq:model_lagrange_multiplier_floweq}. Before we address the resulting active surface Beris-Edwrds models we analyse these forces. 

\begin{figure}[htp]
	\centering
	\begin{tikzpicture}[node distance=0pt]
		\node(Hill) 		at (0,0) 				{\includegraphics[width=0.35\textwidth]{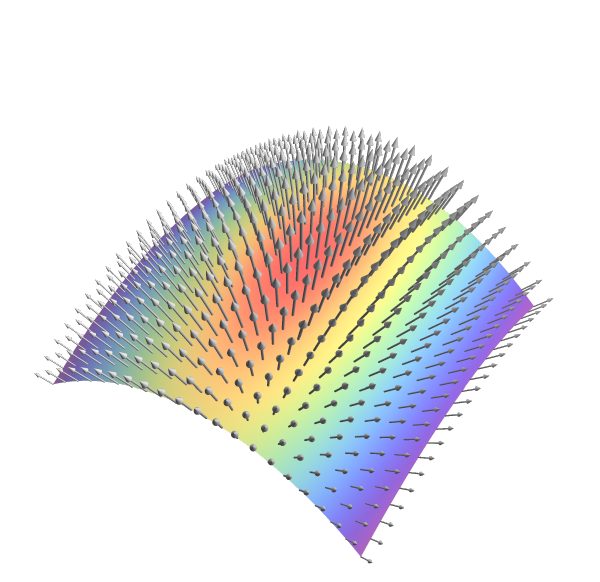}};
		\node(Hillbar) 		[right=of Hill] 		{\includegraphics[width=0.08\textwidth]{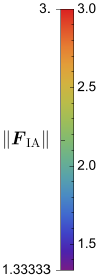}};
		\node(Sattle) 		[right=20pt of Hillbar]	{\includegraphics[width=0.35\textwidth]{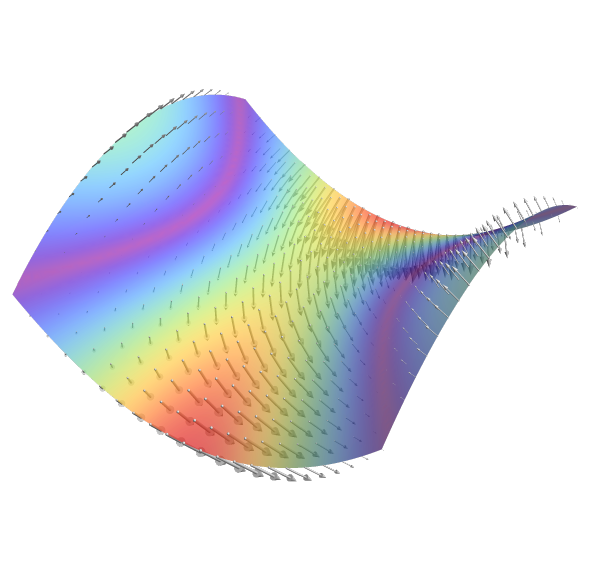}};
		\node(Sattlebar)	[right=of Sattle] 		{\includegraphics[width=0.097\textwidth]{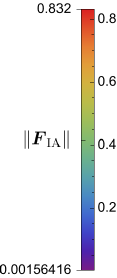}};
	\end{tikzpicture}
	\caption{Active isotropic (geometric) force fields on a dome-shaped (left) and a saddle surface (right) for an active isotropic coefficient $ \cIA=\cAC=-1 $.
		These forces act purely in normal direction and their magnitude is entirely determined by the mean curvature, to be more precise by $ \|\FbIA\| = |\meanc| $.}
	\label{fig:isotropic_activity}
\end{figure}

Figure \ref{fig:isotropic_activity} depicts the active isotropic force field on different surfaces.
This force is tied to surface tension, influencing the energy reaction to changes in area \cite{Al-IzziAlexander_PRR_2023,Capovilla_2002}. Within various approaches this active isotropic force has been used to model mechanical feedback by considering $\cIA$ as a function of a stress regulator molecule, see \cite{bois2011pattern} for the model in one dimension and \cite{mietke2019minimal,wittwer2023computational} for realizations in an axisymmetric setting. However, these approaches consider a compressible surface. Under inextensibility $ \DivC\Vb = 0 $ the active geometric force $ \FbIA $ has no effect on the solution. 


\begin{figure}[htp]
	\centering
	\begin{tikzpicture}[node distance={0pt and 30pt}] 
		\node(Q) 	at (0,0) 		{\includegraphics[width=0.4\textwidth]{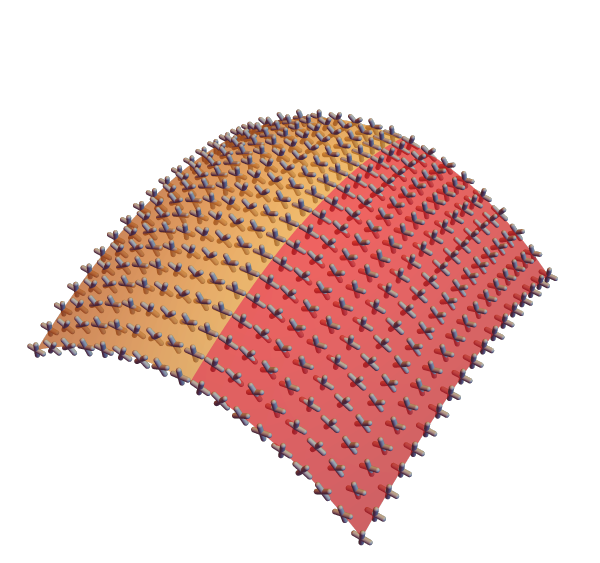}};
		\node(QSC) 	[below=of Q] 	{\includegraphics[width=0.4\textwidth]{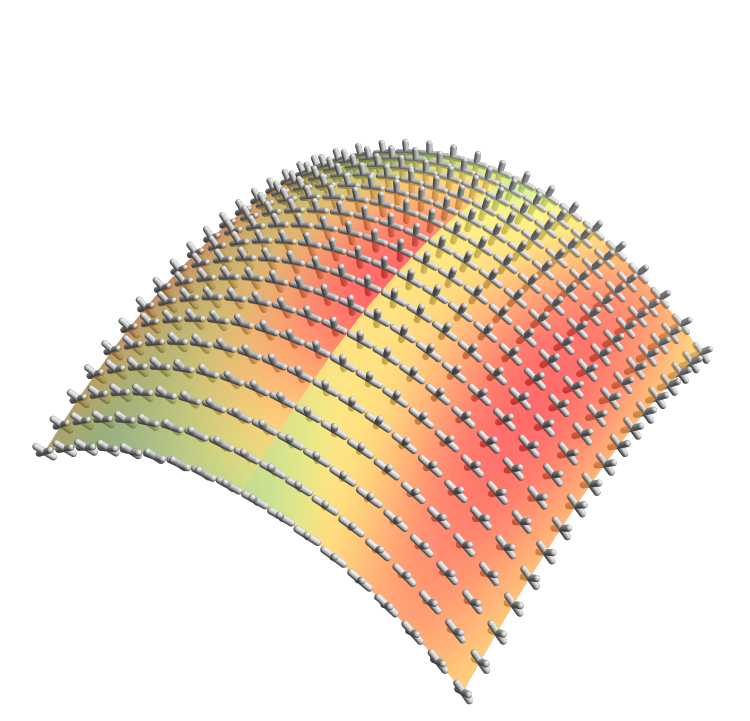}};
		\node(F) 	[right=of Q] 	{\includegraphics[width=0.4\textwidth]{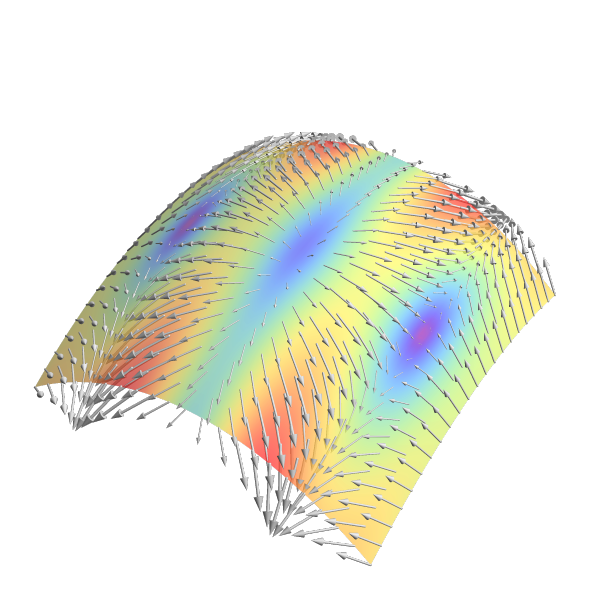}};
		\node(fnor)	[right=of QSC]	{\includegraphics[width=0.4\textwidth]{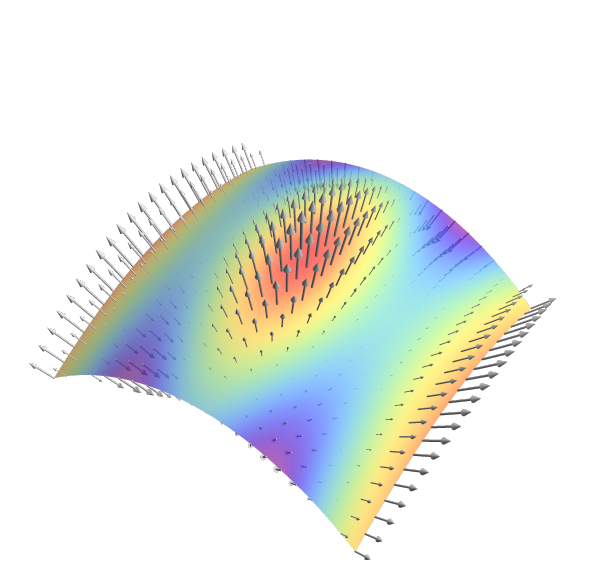}};
		
		\node(orderbar)	[below=of QSC] 	{\includegraphics[width=0.3\textwidth]{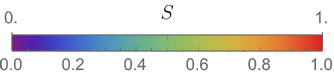}};
		\node(Fbar) 	[below=of F] 	{\includegraphics[width=0.3\textwidth]{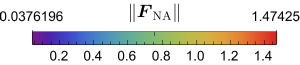}};
		\node(Fnorbar) 	[below=of fnor]	{\includegraphics[width=0.3\textwidth]{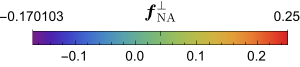}};
		
		\coordinate(QS1) at ($(Q.center)!0.5!(Q.north west)$);
		\node at (QS1) {$S_1$};
		\coordinate(QS2) at ($(Q.center)!0.55!(Q.south east)$);
		\node at (QS2) {$S_2$};
		\coordinate(QSCS1) at ($(QSC.center)!0.85!(QSC.west)$);
		\node at (QSCS1) {$S_1$};
		\coordinate(QSCS2) at ($(QSC.center)!0.55!(QSC.south east)$);
		\node at (QSCS2) {$S_2$};
		
		\coordinate(QMapAnker) 		at ($(Q.center)!0.5!(Q.south west)$); 
		\coordinate(QSCMapAnker) 	at ($(QSC.center)!0.5!(QSC.north west)$); 
		\draw[|->, thick, bend right=30] (QMapAnker) to node[right] {$ \proj_{\tangentCQR}$} (QSCMapAnker);
	\end{tikzpicture}
\caption{Biaxial nematic field (top left), its surface conforming part (bottom left), active nematic force field (top right), and its normal part (bottom right).
		The nematic field is given by the Q-tensor field $ \Qb = \sum_{\alpha=1}^2 S_\alpha(\pb_\alpha\otimes\pb_\alpha - \frac{1}{3}\Id_{\R^3})  $
		for normalized and mutual orthogonal director fields $ \pb_\alpha\in\tangentR $, \ie\ $ \pb_\alpha\pb_\beta = \delta_{\alpha\beta} $.
		Both rods of the crosses depict these directors apolarly.
		The associated scalar order fields are constant with $ S_1\equiv\frac{3}{4} $ and $ S_2 \equiv 1 $.
		They are plotted only on one of the mirrow-symmetric halves of the surface each and could accordingly be symmetrically continued onto their opposite half.
		The active nematic force field $ \FbNA $ does not depend on non-conforming parts of the nematic field.
		As a consequence the surface conforming field $ \proj_{\tangentCQR}\Qb $ yields the same force field as $ \Qb $.}
\label{fig:nematic_activity_biaxial_hill}
\end{figure}

We next consider the active nematic force $ \FbNA $. Figure \ref{fig:nematic_activity_biaxial_hill} yields an example for a biaxial field in $ \tangentQR $. The active nematic force only depends on the conforming parts of the nematic field and contains tangential and normal components. The normal forces substantially scale with the curvature. This provides an additional coupling mechanism between the geometry, the flow field and the Q-tensor field. It even suggests the possibility of an active self-amplifying dynamic effect, potentially leading to buckling. Such behaviour could be found in corrugated sheets forming microtubule/kinesin nematic fluids, see \cite{Senoussi_2019, Str_bing_2020}. In this context again $\cNA$ is considered as a function of the motor concentration. 



Any Q-tensor field $ \Qb\in\tangentQR  $ is uniquely orthogonal decomposable into a tangential (flat degenerated) Q-tensor field $ \qb\in\tangentQS $, 
a scalar-valued normal eigenvector field $ \beta\in\tangentS[^0] $,
and a tangential vector-valued surface non-conforming field $ \etab\in\tangentS $, by
\begin{align}
\Qb &= \qb + \etab\otimes\normal + \normal\otimes\etab + \beta\left( \normal\otimes\normal -\frac{1}{2}\IdS \right) \formComma
	\label{eq:qtensor_decomposition}
\end{align}
see \cite{NitschkeVoigt_2023}.
Substituting this decomposition into the active nematic force field  \eqref{eq:active_nematic_force}
and the tangential part $\IdS\Qb\IdS = \qb - \frac{\beta}{2}\IdS $ of $\Qb$ yields
\begin{align}
    \FbNA  \label{eq:active_nematic_force_conforming}
	&= \cNA\left( \DivC\qb - \frac{1}{2}\GradC\beta \right)
	   = \cNA\left(\div\qb - \frac{1}{2}\nabla\beta + \left( \qb\dbdot\shop - \frac{\meanc}{2}\beta \right)\normal\right) \formPeriod
\end{align}
This analytically confirms the result depicted in Figure \ref{fig:nematic_activity_biaxial_hill}. The force does not depend on any surface non-conforming parts of the Q-tensor field a priori. 
For an inextensible fluid only the flat-degenerated part  $ \qb $ is relevant, since the normal part given by $ \beta $ yields only a pressure-like force and could be translated into a generalized pressure term in the usual way by defining $\tilde{p} = p + \frac{\cNA}{2} \beta$ and considering $\tilde{p}$ in Eq. \eqref{eq:model_lagrange_multiplier_floweq} or the corresponding equations in the models defined below.

\subsection{Energy Rate}\label{sec:energy_rate}

Following \cite{Nitschke2023}, the influence of an energy flux potential $ \fluxpotential_{\alpha} $ on the total energy rate is entirely determined by its acting generalized material forces
$ \Fb_{\alpha}\in\tangentR $ and $ \Hb_{\alpha}\in\tangentQR $, and the process variables $ \Vb\in\tangentR $ and $ \Dmat\Qb\in\tangentQR $, 
through the quantity $ \dot{\energy}_{\alpha}:= \innerH{\tangentR}{\Fb_{\alpha}, \Vb} + \innerH{\tangentQR}{\Hb_{\alpha}, \Dmat\Qb} $.
In absence of a Q-tensor force $ \HbAC $ for activity, the fluid forces Eqs. \eqref{eq:active_geometric_force} and \eqref{eq:active_nematic_force} yield
\begin{align*}
    \dot{\energy}_{\AC} 
        &=  \innerH{\tangentR}{\FbIA + \FbNA, \Vb}
         =  -\innerH{\tangentR[^2]}{\cIA\IdS + \cNA\IdS\Qb, \nablaC\Vb}
         = - \energyAC \formPeriod
\end{align*}
As a consequence the rate of the total energy $ \energyTOT = \energyK + \energyEL + \energyTH + \energyBE $, 
comprising the kinetic and all considered potential energies listed in Section \ref{sec:energetic_contributions}, 
is given by 
\begin{align}\label{eq:energy_rate}
    \ddt\energyTOT
            &= -2\left(  \energyIM + \energyNV \right) -  \energyAC \formPeriod
\end{align}
Since the active energy flux $ \energyAC $, Eqs. \eqref{eq:fluxAC_origin}/\eqref{eq:fluxAC}, is not bounded from below, the resulting active models do not have to be dissipative.
We can only ensure that $ \ddt\energyTOT + \energyAC \le 0 $ holds.

\subsection{Active Surface Beris-Edwards Models}\label{sec:active_beris_edwards_model}

\subsubsection{Active General Surface Q-tensor Models}

Considering the active (isotropic) geometric fluid force \eqref{eq:active_geometric_force} and the active (anisotropic) nematic fluid force \eqref{eq:active_nematic_force} in the (passive) surface Beris-Edwards model leads to the active surface Beris-Edwards model, which reads:

\textit{Find the material velocity field $ \Vb\in\tangentR $, Q-tensor field $ \Qb\in\tangentQR $,  surface pressure field $ p\in\tangentS $ and Lagrange parameter fields $\Lambdab_{\gamma}\in\Vcal_{\gamma}\le\tangentR[^{n_{\gamma}}]$ for all $ \gamma\in\Cset$ \soth}
\begin{subequations}\label{eq:model_lagrange_multiplier_a}
\begin{gather}
    \rho\Dmat\Vb \label{eq:model_lagrange_multiplier_floweq_a}
			=  \GradC\left(\pTH - p\right) 
				+ \left(\fnorBE + \cIA\meanc\right)\normal
                + \DivC\left( \widetilde{\Sigmab} + \cNA\IdS\Qb\IdS \right)
				+ \sum_{\gamma\in\Cset} \Fb_{\gamma} \formComma \\
    \widetilde{M} \Dt[\Phi]\Qb \label{eq:model_lagrange_multiplier_moleculareq_a}
                 = \HbEL + \HbTH + \widetilde{\Hb}^{\Phi}_{\NV} + \sum_{\gamma\in\Cset} \Hb_{\gamma} \formComma\\
    0 = \DivC\Vb \formComma\quad\quad
    \nullb = \Cb_{\gamma}, \quad \forall\gamma\in\Cset \label{eq:model_lagrange_multiplier_constraineq_a}
\end{gather}
\end{subequations}
\textit{holds for $ \dot{\rho}=0 $ and given initial conditions for $ \Vb $, $ \Qb $ and mass density $ \rho\in\tangentS[^0] $.}

The passive stress field $\widetilde{\Sigmab}$, the nematic viscosity Q-tensor force 
$ \widetilde{\Hb}^{\Phi}_{\NV}$ and the immobility coefficient $ \widetilde{M}$ are defined as in the (passive) surface Beris-Edwards model, also all mandatory quantities are still given in Table \ref{tab:forces} and optional constraint quantities in Table \ref{tab:constraints} within \ref{app:a}. 
Again the preferred immobility mechanisms is determined by the symbol $ \Phi\in\{\jau,\mfrak\} $ and also the restrictions and abbreviations remain. 

Even if this general formulation is suggested in \cite{Nitschke2023} also for the case of the Q-tensor field to be surface conforming, in the following we also formulate an alternative model with this constraint enforced directly, as it provides additional physical insight.

\subsubsection{Active Surface Conforming Q-tensor Models}

To enforce $ \Qb\in\tangentCQR $ we consider the ansatz \eqref{eq:qtensor_decomposition} for a vanishing non-conforming component $ \etab=\nullb $.
This yields $ \Qb = \qb + \beta(\normal\otimes\normal - \frac{1}{2}\IdS)$, 
where $ \qb\in\tangentQS $ and $ \beta\in\tangentS[^0] $ are the remaining mutual independent degrees of freedoms determining the surface conforming nematic field.
Similar to the general model, we only add the active forces  \eqref{eq:active_geometric_force} (geometric/isotropic) 
and \eqref{eq:active_nematic_force_conforming} (nematic/anisotropic) to the passive surface conforming model in \cite{Nitschke2023}.
With an orthogonal splitting \wrt\ tangential and normal spaces, the active surface conforming Beris-Edwards model reads:

\textit{Find the tangential and normal material velocity fields $ \vb\in\tangentS $ and $ \vnor\in\tangentS[^0] $, tangential Q-tensor field $ \qb\in\tangentQS $, normal eigenvalue field $ \beta\in\tangentS[^0] $, pressure field $ p\in\tangentS[^0] $ and Lagrange parameter fields $ \lambdab_{\gamma}\in\Vcal_{\gamma} $ for all $ \gamma\in\Cset_{\SC} $ \soth}
\begin{subequations}\label{eq:model_conforming}
   \begin{gather}
   \begin{align}
        \rho\ab \label{eq:model_conforming_tangentialeq}
            &= \nabla\left(\pTH - p - \frac{\cNA}{2}\beta\right)   + \div\left(  \widetilde{\sigmab} + \cNA\qb\right) 
                    -\shop\zetabSC + \sum_{\gamma\in\Cset_{\SC}}\fb_{\gamma}\formComma\\
        \rho\anor \label{eq:model_conforming_normalseq}
            &= \left(\pTH - p + \cIA - \frac{\cNA}{2}\beta\right)\meanc + \fnorBE 
                +\shop\dbdot\left( \widetilde{\sigmab} + \cNA\qb \right)
             + \div\zetabSC + \sum_{\gamma\in\Cset_{\SC}} \fnor[\gamma] \formComma
   \end{align}\\
   \begin{align}
        \widetilde{M}\timeD\qb \label{eq:model_conforming_qtensor}
            &= \hbEL + \hbTH + \widetilde{\hb}_{\NV}^{\Phi} + \sum_{\gamma\in\Cset_{\SC}} \hb_{\gamma} \formComma \\
        \widetilde{M} \dot{\beta} \label{eq:model_conforming_betaeq}
            &= \omegaEL + \omegaTH  + \widetilde{\omega}_{\NV} + \sum_{\gamma\in\Cset_{\SC}} \omega_{\gamma}\formComma
   \end{align}\\
   \div\vb = \vnor\meanc \formComma \quad
   \nullb = \Cb_{\gamma} \formComma \quad 
   \forall\gamma\in\Cset_{\SC} \label{eq:model_conforming_constraineqs}
   \end{gather}
\end{subequations}
\textit{holds for $ \dot{\rho}=0 $ and given initial conditions for $ \vb $, $ \vnor $, $ \qb $, $ \beta $ and mass density $ \rho\in\tangentS[^0] $.}

For the representations of the tangential and normal acceleration, $ \ab $ and $ \anor $, see Table \ref{tab:forces_conforming} in \ref{app:a}. The passive stress field is summarized to
\begin{align*}
    \widetilde{\sigmab} 
        &= \sigmabEL + \sigmabIM + \sigmabNV^0 + \xi\sigmabNV^1 +  \xi^2\sigmabNV^2 \formComma
\end{align*}
and the nematic viscosity Q-tensor force partly to 
\begin{align*}
    \widetilde{\hb}_{\NV}^{\Phi} &= \xi(\hbNV^1 + \xi\widetilde{\hb}^{2,\Phi}_{\NV}) 
    &&\text{and}
    &\widetilde{\omega}_{\NV} &= \xi^2\widetilde{\omega}^{2}_{\NV} \formPeriod
\end{align*}
The effective part of the surface conforming constrain stress $ \normal\otimes\zetabSC $ is given by $ \zetabSC = (3\beta\IdS - 2\qb)(\zetabEL + \zetabIM) $.
To determine all terms sufficiently, mandatory quantities are given in Table \ref{tab:forces_conforming} and optional constraint quantities, such as $ \fb_{\gamma}, \fnor[\gamma], \hb_{\gamma}, \omega_{\gamma}, \Cb_{\gamma}$ and $\Vcal_{\gamma}$,  are given in Table \ref{tab:constraints_conforming} within \ref{app:a}.
These constraints can be chosen by stipulating the constrain identifier set $ \Cset_{\SC}\subset\{ \CB, \UN, \NN, \NF\} $.  
Note that the surface conforming model \eqref{eq:model_conforming} is equivalent to the general model \eqref{eq:model_lagrange_multiplier_a} with $ \SC\in\Cset $.
Beside the transition to the tangential calculus, we only determined the surface conforming Lagrange-parameter $ \lambdabSC $ and substituted it into the general model as shown in \cite{Nitschke2023}.

\subsection{Comparison with Existing Models} \label{sec:existing}

Various special cases of the derived active surface Beris-Edwards models can be related to previously proposed models. In the following we consider three examples.

\subsubsection{Self-Deforming Active Nematic Shells} \label{sec:selfdefshell}

Considering the surface conforming constraint $\SC \in \Cset$ in Eqs. \eqref{eq:model_lagrange_multiplier_a} 
the model can be compared with \cite{Metselaar_2019}. However, a one-to-one comparison of the terms is not possible. Instead of an explicit formulation as an active surface Beris-Edwards model the approach in \cite{Metselaar_2019} considers the surface as a diffuse interface between two isotropic bulk phases and only a three dimensional model for the whole system is proposed. The nematodynamic model at the interface is enforced by an isotropic-nematic phase transition and the surface conforming constraint is realized by an interfacial anchoring free energy term. In principle, letting the diffuse interface width go to zero, formal matched asymptotic expansions allow to derive the corresponding sharp interface model. For a general discussion of approximating surface partial differential equations by diffuse interface approaches see \cite{Raetz_2006,Nestler_2024}. Especially for tangential vector- or tensor-fields this comes with various technical subtleties and convergence of the diffuse interface model to the sharp model requires higher order approximations of the involved geometric quantities \cite{Nestler_2024}. However, such an expansion is not performed and thus the terms cannot be compared. One difference, which becomes apparent already in the three dimensional formulation is the mechanism for ``flow alignment''. Instead of anisotropic viscosity a linear reaction term is considered, see Appendix \ref{sec:flow_alignment} for a comparison of these approaches. If the model in \cite{Metselaar_2019} is uniaxial can not be judged. A Jaumann ($ \Phi=\jau $) model is considered, the same active nematic force field is used and also the other terms seem to be consistent. 

\subsubsection{Anisotropic Linearized Surface-Conforming Eulerian-Jaumann Model} \label{sec:anisolinmod}

Considering a geometrically stationary surface strongly simplifies the model but also leads to previously discussed approaches. Treating the surface conforming Jaumann ($ \Phi=\jau $) model \eqref{eq:model_conforming} in the same way as in \cite{Nitschke2023} to elaborate the expansion, \wrt\ the viscous anisotropy coefficient $ \xi $, 
up to linear order, considering the stationary surface constraint ($ \NN $), \ie\ $ \Vb=\vb\in\tangentS $, restricting to a constant eigenvalue in normal direction ($ \CB $), \ie\ $ \Qb=\qb + \beta(\normal\otimes\normal - \frac{1}{2}\IdS) \in\tangentCQR $ with $ \beta\in\R $, and considering an Eulerian perspective, \ie\ $ \Vb_{\!\ofrak}=\nullb $, yield 
\begin{subequations}\label{eq:model_linearized_jaumann_conforming}
	\begin{gather}
		\begin{align}
			\rho\left( \partial_t \vb + \nabla_{\vb}\vb \right) 
				&= -\nabla p 
					+ \div\left(\coeffIF\left(\nabla\vb + (\nabla\vb)^T\right) + \cNA\qb + \qb\hb_\potenergy - \hb_\potenergy \qb -\frac{\coeffIF\xi}{M}\hb_\potenergy - \coeffIF\xi\hat{\sigmab} \right)
					 -\hb_\potenergy\dbdot\nabla\qb + \landau(\xi^2) \formComma
		\end{align}\\
		\begin{align}
			\partial_t \qb + \nabla_{\vb}\qb - \frac{1}{2}\left(\nabla\vb - (\nabla\vb)^T\right)\qb + \frac{1}{2}\qb\left(\nabla\vb - (\nabla\vb)^T\right)
			&=  \frac{1}{M}\hb_\potenergy + \frac{\coeffIF\xi}{2M}\left(\nabla\vb + (\nabla\vb)^T\right) + \landau(\xi^2) \formComma 
		\end{align}\\
		\div\vb
			= 0 \formComma
	\end{gather}
\end{subequations}
where
\begin{align*}
    \hb_\potenergy 
        &=  L \left(  \Delta\qb - (\meanc^2-2\gaussc)\qb + 3\beta \meanc \left( \shop - \frac{\meanc}{2}\IdS \right) \right)
        	-\left(2a - 2b\beta + 3c\beta^2  + 2c\Tr\qb^2\right)\qb \formComma\\
    \hat{\sigmab}
        &= \qb\left(\nabla\vb + (\nabla\vb)^T\right) + \left(\nabla\vb + (\nabla\vb)^T\right)\qb - \beta\left(\nabla\vb + (\nabla\vb)^T\right) \formComma\\
    -\hb_\potenergy\dbdot\nabla\qb
                &= \nabla\pTH + \div\sigmabEL + \left(2\shop\qb - 3\beta\shop \right)\zetabEL \formPeriod \notag 
\end{align*}
This model is comparable with \cite{Nestler_2022}. The differences lie in a linear flow alignment mechanism instead of nematic viscosity, \cf\ Appendix~\ref{sec:flow_alignment},
an additional external friction term, and absence of the Ericksen stress. The argument which allows neglecting the Ericksen stress, originates from \cite{Giomi_2015}. It is based on the assumption that this stress comprises only higher orders in the derivatives of $ \qb $ on a flat surface.
However, on a curved surface also lower order effects in  $ -\hb_\potenergy\dbdot\nabla\qb $ are present, 
\eg\ due to the structure of covariant derivatives or the presence of the surface-conforming constraint force $ \left(2\shop\qb - 3\beta\shop \right)\zetabEL  $. This makes neglecting the Ericksen stress on curved surfacs questionable. However, other proposed models consider even stronger simplifications. The approach in \cite{Pearce_2019} considers $ \beta=0 $ and nematic coupling to the fluid equation is almost completely avoided up to activity.

Considering Eqs. \eqref{eq:model_linearized_jaumann_conforming} on a flat surface leads to even more drastic simplifications. We obtain $\beta=0$ and $\shop=0$. The molecular field and the consequent force of the Ericksen stress read
\begin{align*}
	\hb_\potenergy 
	&=  L \Delta\qb - 2\left(a  + c\Tr\qb^2\right)\qb \formComma
	&-\hb_\potenergy\dbdot\nabla\qb
	&= \nabla\pTH + \div\sigmabEL \formPeriod
\end{align*}
The resulting model relates to classical active nematodynamic models in two dimensions, see \cite{marchetti2013hydrodynamics, Doostmohammadi_2018} for reviews. Differences between these models are mainly due to the realization of "flow alignment". In \cite{Giomi_2014,Giomi_2015,Juelicher_2018,Mirza2023} it is implemented as a linear reaction, and in 
\cite{Thampi_2016,Rorai_2021} using an extended corotation. While common in the classical liquid crystal theory \cite{Leslie_1966,Beris_1994,Parodi_1970}, we are not aware of an active model considering anisotropic viscosity. For a comparison of these approaches we refer to Appendix \ref{sec:flow_alignment}.

\subsubsection{Effective Active Surface Models} \label{sec:effactmodd}

\begin{figure}[htp]
\centering
\begin{tikzpicture}[node distance=0pt]
    \node(E) 	at (0,0)		{\includegraphics[width=0.3\textwidth]{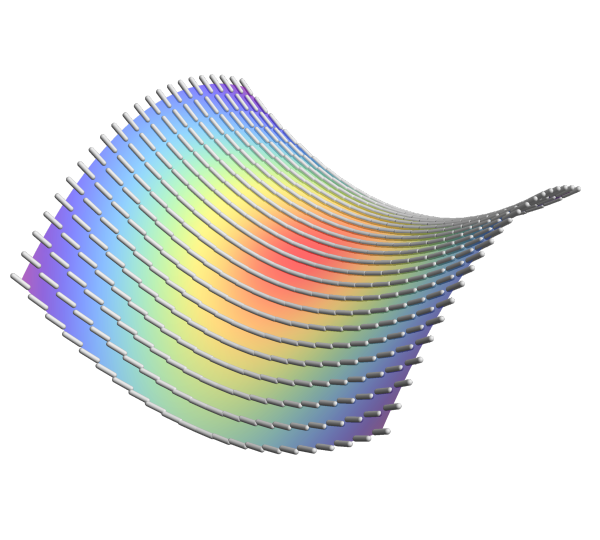}};
    \node(FE)	[right=of E]	{\includegraphics[width=0.3\textwidth]{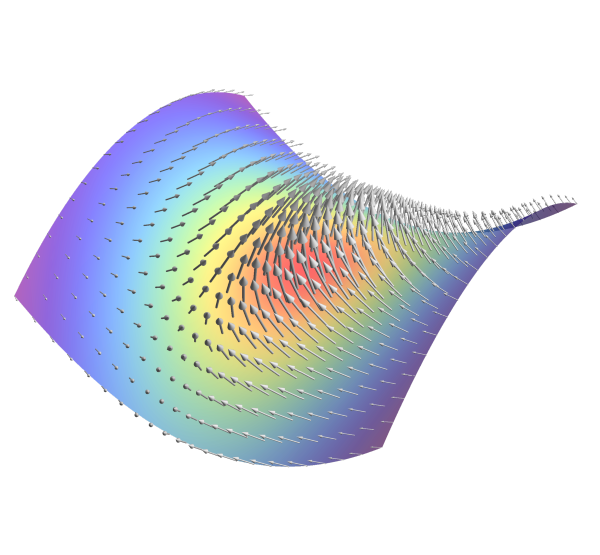}};
    \node(NFE)	[right=of FE]	{\includegraphics[width=0.3\textwidth]{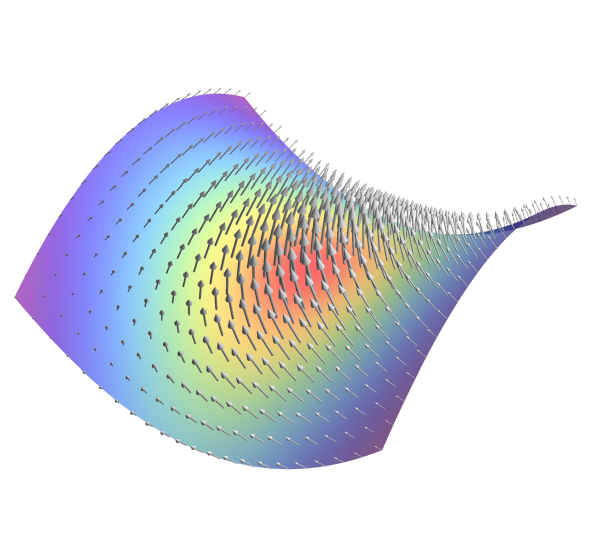}};
    
    \node(EBar)		[below=-15pt of E]	{\includegraphics[width=0.3\textwidth]{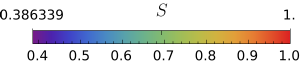}};
    \node(FEBar)	[right=of EBar]		{\includegraphics[width=0.3\textwidth]{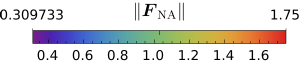}};
    \node(NEFBar)	[right=of FEBar]	{\includegraphics[width=0.3\textwidth]{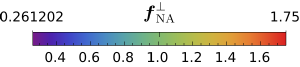}};
    
    \node(TE) [left=+10pt of E, rotate=90] {even};
    
    \node(O) 	[below=of EBar]		{\includegraphics[width=0.3\textwidth]{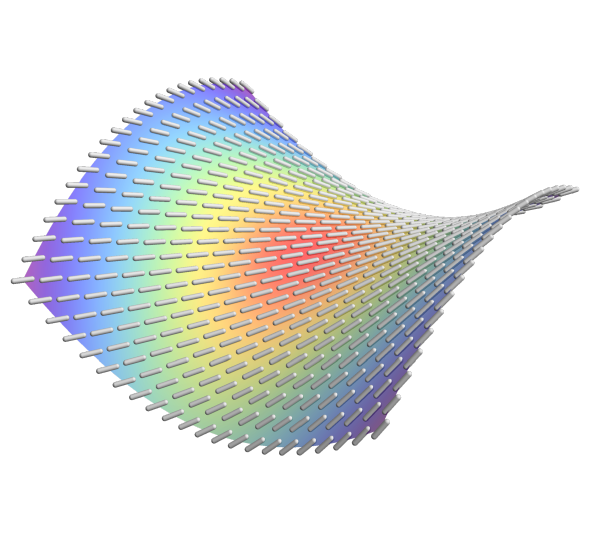}};
    \node(FO)	[right=of O]		{\includegraphics[width=0.32\textwidth]{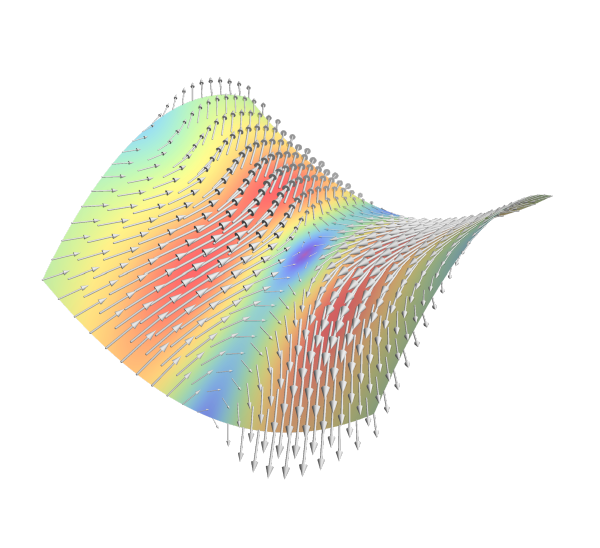}};
    \node(NFO)	[right=of FO]		{\includegraphics[width=0.3\textwidth]{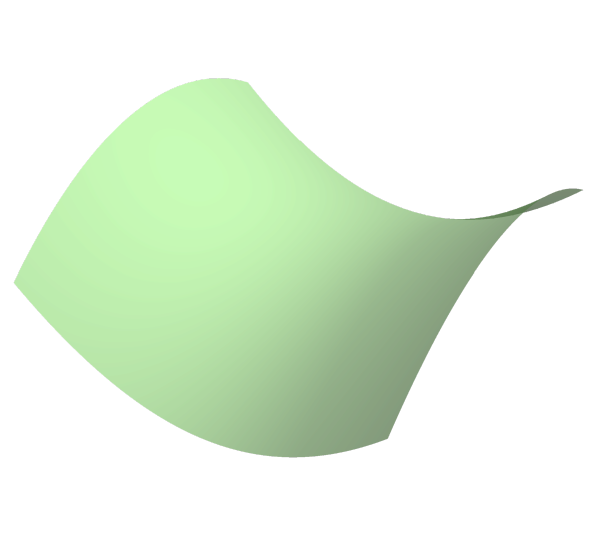}};
    
    \node(OBar)		[below=-15pt of O]	{\includegraphics[width=0.3\textwidth]{IMG/saddle_Geo_bebo2db7_degenerated_nemates_order_barlegend.pdf}};
    \node(FOBar)	[right=of OBar]		{\includegraphics[width=0.3\textwidth]{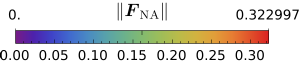}};
    \node(NFOBar)	[right=of FOBar, yshift=4pt]		{\includegraphics[width=0.28\textwidth]{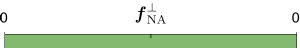}};
    
    \node(TO) [left=+10pt of O, rotate=90] {odd};
    
\end{tikzpicture}
\caption{Depictions of the geometric ansatz \eqref{eq:geometric_ansatz} as nematic fields (left), the resulting anistropic active forces (middle) and their normal components (right).
	We consider a pure even-ansatz (top) with $ (\ceven,\codd) = (\frac{2}{7}, 0) $ and a pure odd-ansatz (bottom) with $ (\ceven,\codd) = (0, \frac{2}{7}) $.
	The normal component for the odd-activity (bottom right) is always vanishing and shown here for completeness only.}
\label{fig:evenodd}
\end{figure}

Considering the surface conforming and constant-$ \beta $ constraint $\SC,\CB \in \Cset$ with $ \beta=0 $, leads to $\Qb = \qb$. If we in addition assume that $\Qb$ instantaneously responds to the surface, such as formally setting $\widetilde{M} = \infty$, allows to approach the active nematic force $\FbNA$ with pure geometric terms. In order to achieve this we consider the ansatz 
\begin{align}
	\Qb \label{eq:geometric_ansatz}
		&= \ceven\shopQ +\codd\hodge\!\shopQ
\end{align}
with $\shopQ = \shop - \frac{\meanc}{2}\IdS$, its orthogonal complement $\hodge \shopQ = - \Eb\shop + \frac{\meanc}{2}\Eb$, and for some coefficients $ \ceven,\codd\in\R $.
Following \cite{Al-IzziAlexander_PRR_2023}, we refer terms including $ \ceven $ as ``even'' and terms including $ \codd $ as ``odd'' components. 
This characterizes the number of repeated applications of the Hodge operator $*$, which describes a complex structure through $*\circ* = *^2 = -\IdS$.
Note that $\shopQ, \hodge\shopQ \in\tangentQS$ are forming an orthogonal system of equal length, 
\ie\ $ \shopQ\dbdot\hodge\shopQ = 0 $ and $ \|\shopQ\|^2 =  \|\hodge\shopQ\|^2 $. Therefore, if and only if  $\| \shop - \frac{\meanc}{2}\IdS \|^2 = \meanc^2 - 4\gaussc$ is nowhere vanishing, they provide a 2-dimensional $ \R $-vector space within the space of flat-degenerated Q-tensor fields $ \tangentQS $.
The even-part represents a tangential nematic field along the principal curvature directions, \cf\ Fig.~\ref{fig:evenodd} (top left), thus mimicking the alignment with principle curvature directions resulting from extrinsic curvature contribution in the equation for the Q-tensor field \cite{napoli2012surface,golovaty2017dimension,Nitschke_2018}.  
Formally, let $ \shop\kb_{\{1,2\}} = \kappa_{\{1,2\}}\kb_{\{1,2\}} $, where $ \kappa_{\{1,2\}}\in\tangentS[^0] $ are the principle curvatures 
and $ \kb_{\{1,2\}}\in\tangentS $ their corresponding directions. 
We stipulate $ \|\kb_{\{1,2\}}\| = 1 $ and $ \kb_2 = \hodge\kb_1 = -\Eb\kb_1 $ to close ambiguities in length and orientation.
Using this approach the even-part is representable as 
\begin{align*}
	\ceven\shopQ
		&= \ceven\frac{\kappa_1-\kappa_2}{2}\left( \kb_1\otimes\kb_1 - \kb_2\otimes\kb_2 \right)
		 = \ceven\left( \kappa_1-\kappa_2 \right) \left(  \kb_1\otimes\kb_1 - \frac{1}{2}\IdS \right) \formComma
\end{align*}
\ie\ it holds $ \ceven\shopQ\kb_{\{1,2\}} = \{+,-\} \ceven\frac{\kappa_1-\kappa_2}{2}\kb_{\{1,2\}} $.
Analogously, the odd-part yields
\begin{align*}
	\codd\hodge\shopQ
	&= \codd\frac{\kappa_1-\kappa_2}{2}\left( \halfhodge\kb_1\otimes\halfhodge\kb_1 - \halfhodge\kb_2\otimes\halfhodge\kb_2 \right)
	= \codd\left( \kappa_1-\kappa_2 \right) \left(  \halfhodge\kb_1\otimes\halfhodge\kb_1 - \frac{1}{2}\IdS \right) \formComma
\end{align*}
where $ \halfhodge\kb_{\{1,2\}} = \frac{1}{\sqrt{2}}(\{+,-\}\kb_1 + \kb_2) $ is $ \kb_{\{1,2\}} $ rotated by $ \frac{\pi}{4} $ counterclockwise.
It holds the eigenequations $ \codd\hodge\shopQ\halfhodge\kb_{\{1,2\}} = \{+,-\}\codd\frac{\kappa_1-\kappa_2}{2}\halfhodge\kb_{\{1,2\}} $,
\ie\ the odd-part of \eqref{eq:geometric_ansatz} addresses the directions lying exactly in between both principle curvatures directions, \cf\ Fig.~\ref{fig:evenodd} (bottom left).

Let the ansatz \eqref{eq:geometric_ansatz} be substituted into the active nematic force \eqref{eq:active_nematic_force_conforming} for $ \cNA=1 $ (\oeda).
Hodge-compatibility of the covariant derivative, especially $ \div\circ\hodge = \hodge\circ\div $, orthogonalities $ \hodge\shop,\Eb\bot\shop $, and $ \div\shopQ=\frac{1}{2}\nabla\meanc $
lead to
\begin{align*}
	\FbNA 
		&=  \ceven\DivC\shopQ + \codd\DivC\hodge\shopQ\\
		&= \ceven\div\shopQ + \codd\div\hodge\shopQ + \left(  \ceven\shopQ\dbdot\shop +\codd\hodge\shopQ\dbdot\shop \right)\normal\\
		&= \frac{\ceven}{2}\nabla\meanc +  \frac{\codd}{2}\Rot\meanc + \frac{\ceven}{2}\left( \meanc^2 - 4\gaussc \right)\normal \formComma
\end{align*}
where $\Rot\meanc := \hodge\nabla\meanc = -\Eb\nabla\meanc \in\tangentS$. 
Using this in the active surface conforming Q-tensor models \eqref{eq:model_conforming} and considering isotropic viscosity ($ \xi=0 $) results in
\begin{subequations}\label{eq:model_geo_active_decomposed}
\begin{gather}
	\begin{align}
		\rho\ab \label{eq:model_geo_active_tangential}
		&= -\nabla\left( p - \cIA - \frac{\ceven}{2}\meanc \right) + \fnorBE + \frac{\codd}{2}\Rot\meanc +  \div\left( \coeffIF(\nabla\vb + (\nabla\vb)^T - 2\vnor\shop) \right)\formComma\\
		\rho\anor \label{eq:model_geo_active_normal}
		&= -\left( p - \cIA - \frac{\ceven}{2}\meanc \right)\meanc - 2\ceven\gaussc + 2\coeffIF\left(\shop\dbdot\nabla\vb -\vnor(\meanc^2- 2\gaussc) \right) \formComma
	\end{align}\\
	 \div\vb = \vnor\meanc \formPeriod
\end{gather}
\end{subequations}
For the representations of the tangential and normal acceleration, $ \ab $ and $ \anor $, as well as the relevant bending force component $ \fnorBE $, see Table \ref{tab:forces_conforming} in \ref{app:a}. Note that in Eqs. \eqref{eq:model_geo_active_decomposed} the odd active forces act only in tangential and the even active forces only in normal direction. This becomes evident by identifying $\tilde{p} =  p - \cIA - \frac{\ceven}{2}\meanc $ as a generalized surface pressure, which does not impact the velocity field $ \Vb = \vb + \vnor\normal $. The resulting model can be viewed as a fluid deformable surface model \cite{Torres-Sanchez_2019,Reuther_2020,Krause_2023} with even and odd active geometric forces. Figure \ref{fig:evenodd} provides an example of the active nematic force $ \FbNA $, considering the even and odd contributions separably. \\

\noindent {\bf Remark}: Formally, Eqs. \eqref{eq:model_geo_active_decomposed} can also be derived by implementing the ansatz \eqref{eq:geometric_ansatz} as a state constraint in Eqs. \eqref{eq:model_conforming} with the Lagrange-multiplier technique; or by considering the entire model independently of $ \Qb $ in the first place and stipulate the active flux potential \eqref{eq:fluxAC_origin} \wrt\ ansatz \eqref{eq:geometric_ansatz} within the Lagrange-D'Alembert principle. We here refrain from these derivations. \\

Neglecting inertia the effective active surface model \eqref{eq:model_geo_active_decomposed} leads to the model proposed in  \cite{Al-IzziAlexander_PRR_2023}. However, some differences occur. Even forces are neglected in the proposed model in \cite{Al-IzziAlexander_PRR_2023}. Furthermore also additional higher order forces are considered in \cite{Al-IzziAlexander_PRR_2023}. Such higher order forces could be obtained by considering higher powers of curvature terms within the ansatz \eqref{eq:geometric_ansatz} to be used in the anisotropic metric \eqref{eq:nematic_metric}. We here refrain from these derivations. In any case the established connection between the geometric membrane forces that couple elastic membrane bending to in-plane stresses, as introduced in \cite{Al-IzziAlexander_PRR_2023} and the nematic metric \eqref{eq:nematic_metric} provides additional understanding on this effective modeling approach. 

While computationally more feasible than the active nematodynamic models on deformable surfaces, Eqs. \eqref{eq:model_lagrange_multiplier_a} or \eqref{eq:model_conforming}, the validity of the effective active surface model Eqs. \eqref{eq:model_geo_active_decomposed} in this context still has to be confirmed. There are certainly configurations in which the nematic equilibrium can be realized by the ansatz \eqref{eq:geometric_ansatz}. One example is a tube, which is also the configuration analytically explored in \cite{Al-IzziAlexander_PRR_2023}. However, for any closed surface, with the topological requirement of the presence of defects, the ansatz \eqref{eq:geometric_ansatz} can only be a rough approximation, as it does not support defects. 
Moreover, \eqref{eq:model_geo_active_decomposed} implies a break of the up-down symmetry, particularly, the surface orientation determines the sign of the active even force.
Contrarily, the active nematodynamic models, Eqs. \eqref{eq:model_lagrange_multiplier_a} or \eqref{eq:model_conforming}, do not yield such symmetry breaking.
For more general discussions on up-down symmetry breaking formulations we refer to \cite{Salbreux_2022}.

\section{Discussion} \label{sec:diss}

\begin{figure}[htp]
	\centering
	\begin{tikzpicture}[node distance=0pt]
		\node(P12) 	at (0,0) 		{\includegraphics[width=0.24\textwidth]{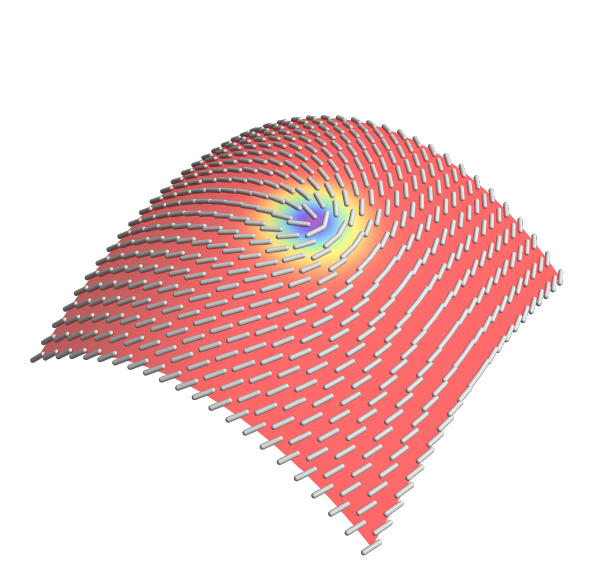}};
		\node(M12) 	[right=of P12] 	{\includegraphics[width=0.24\textwidth]{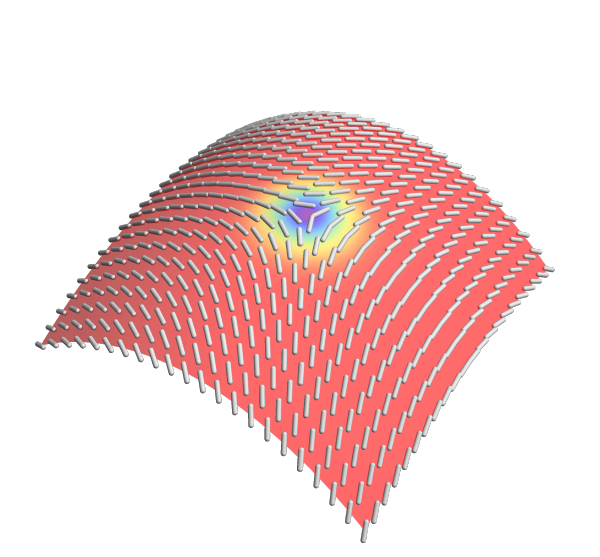}};
		\node(P1) 	[right=of M12]  {\includegraphics[width=0.24\textwidth]{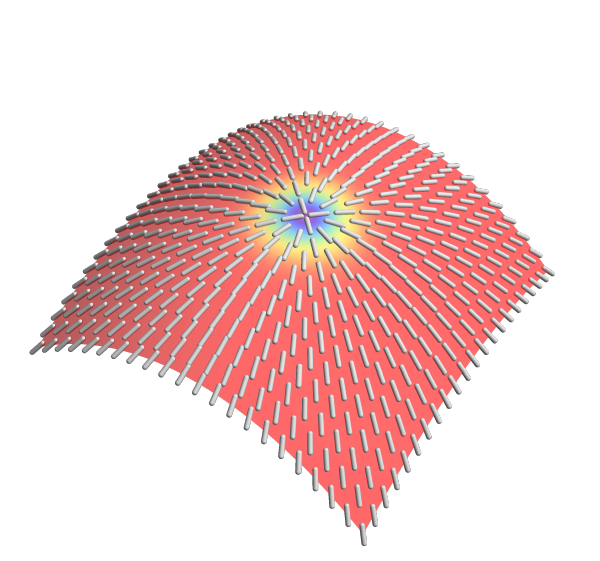}};
		\node(M1) 	[right=of P1]  	{\includegraphics[width=0.24\textwidth]{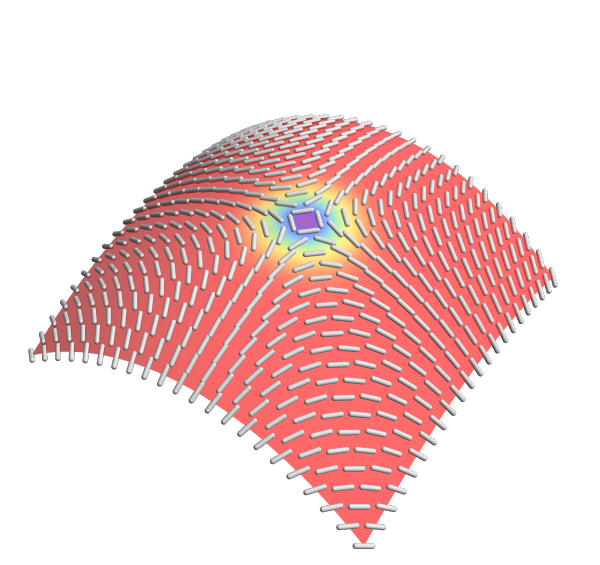}};
		\node(FP12)	[below=20pt of P12]	{\includegraphics[width=0.24\textwidth]{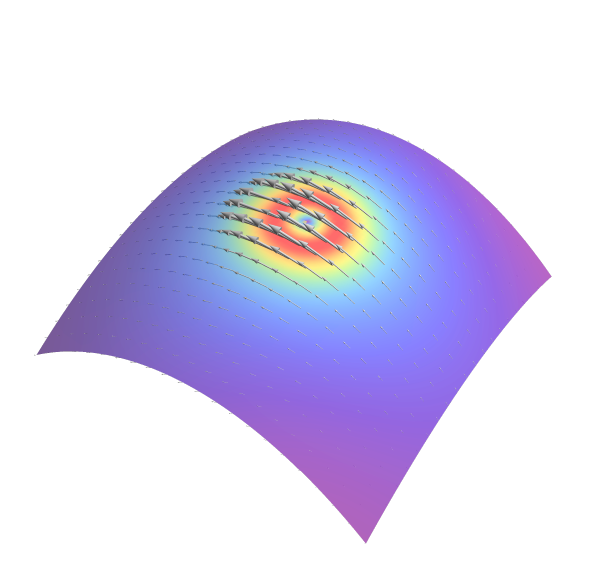}};
		\node(FM12)	[right=of FP12]		{\includegraphics[width=0.24\textwidth]{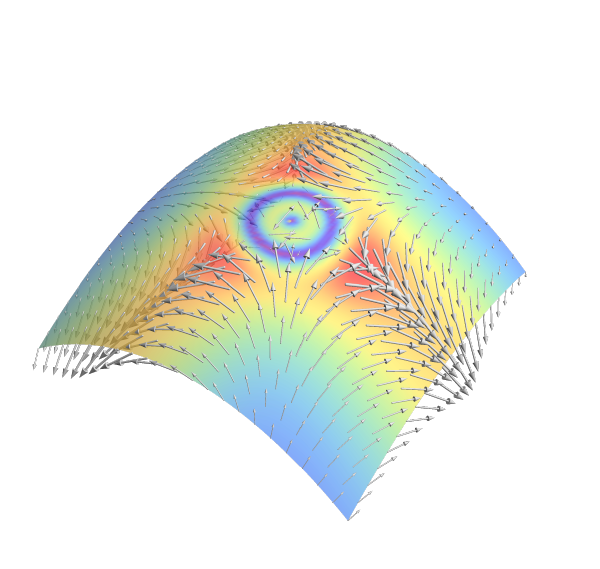}};
		\node(FP1) 	[right=of FM12]		{\includegraphics[width=0.24\textwidth]{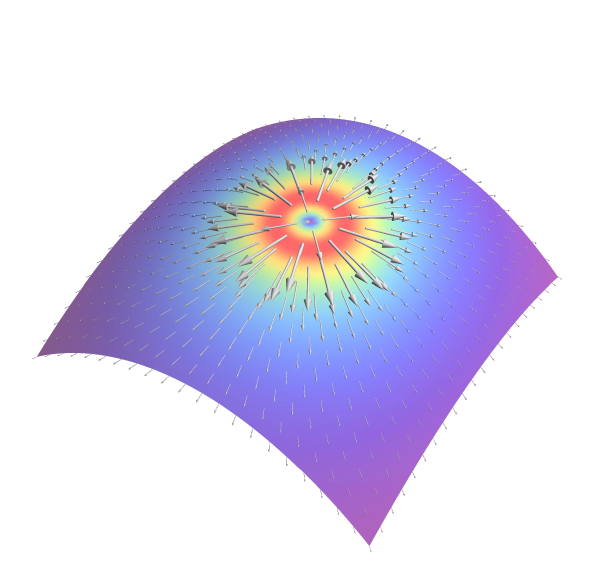}};
		\node(FM1) 	[right=of FP1]  	{\includegraphics[width=0.24\textwidth]{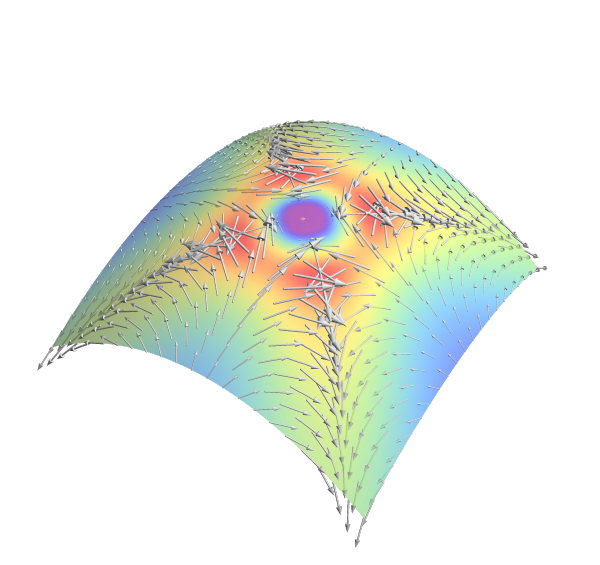}};
		\node(NFP12)	[below=of FP12]		{\includegraphics[width=0.24\textwidth]{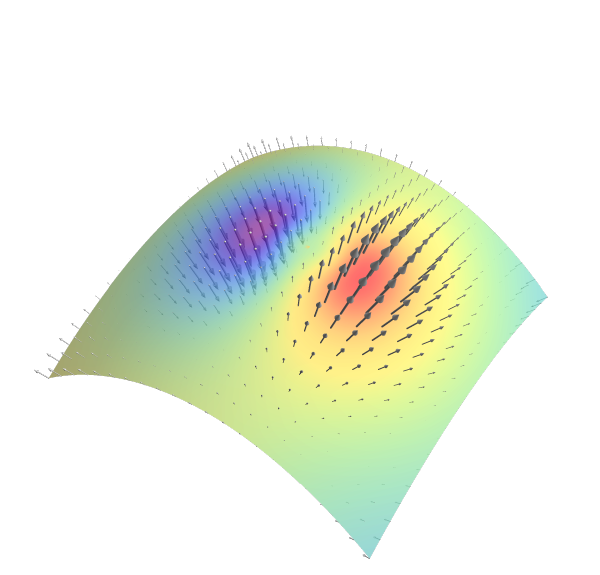}};
		\node(NFM12)	[right=of NFP12]	{\includegraphics[width=0.24\textwidth]{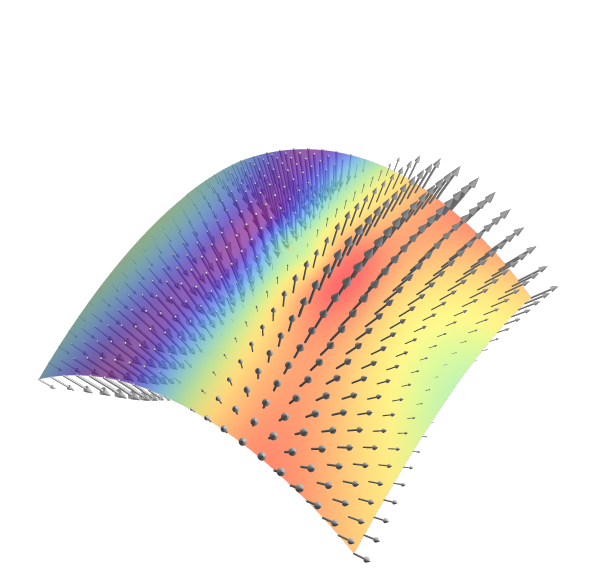}};
		\node(NFP1) 	[right=of NFM12]	{\includegraphics[width=0.24\textwidth]{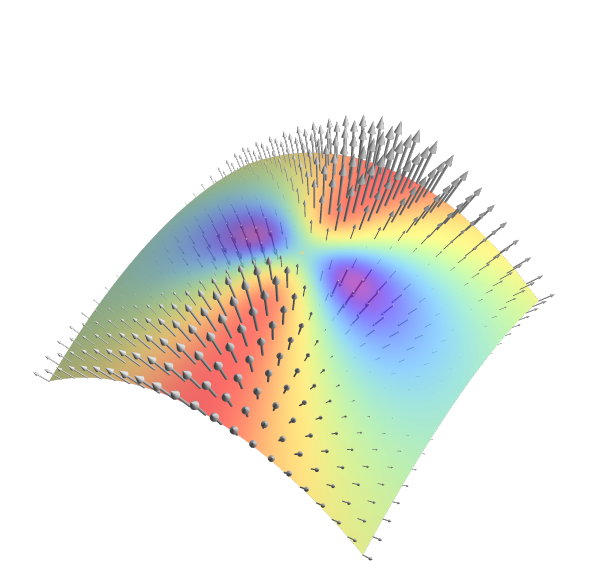}};
		\node(NFM1) 	[right=of NFP1]  	{\includegraphics[width=0.24\textwidth]{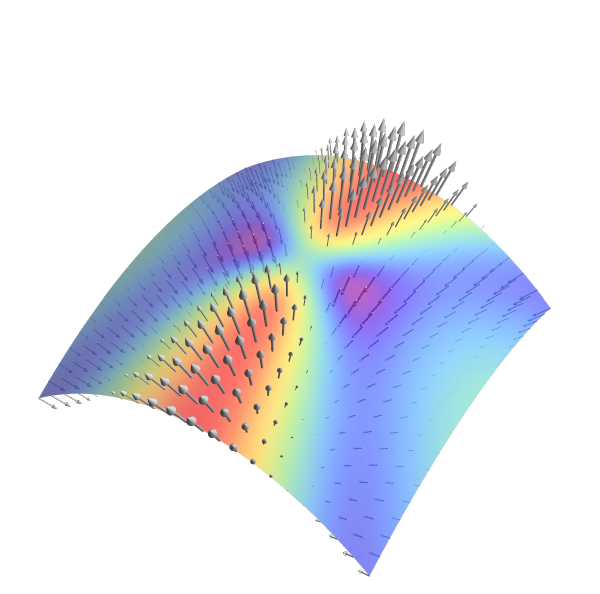}};
		\node(SBar) [below=of P12, xshift=0.12\textwidth] {\includegraphics[width=0.24\textwidth]{IMG/orderbarlegend.pdf}};
		\node(SBarGraph) [below=of P1, xshift=0.10\textwidth] {\includegraphics[width=0.5\textwidth]{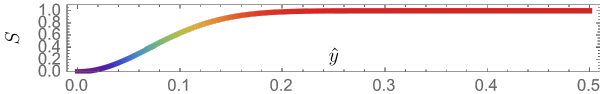}};
		\node(FP12Bar)	[below=-7pt of FP12]	{\includegraphics[width=0.24\textwidth]{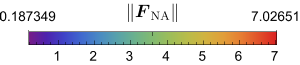}};
		\node(FM12Bar)	[right=of FP12Bar]		{\includegraphics[width=0.24\textwidth]{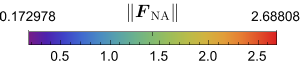}};
		\node(FP1Bar)	[right=of FM12Bar]		{\includegraphics[width=0.24\textwidth]{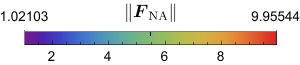}};
		\node(FM1Bar)	[right=of FP1Bar]		{\includegraphics[width=0.24\textwidth]{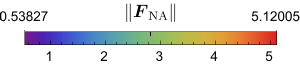}};
		\node(NFP12Bar)	[below=-7pt of NFP12]	{\includegraphics[width=0.24\textwidth]{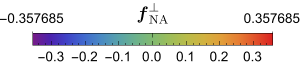}};
		\node(NFM12Bar)	[right=of NFP12Bar]		{\includegraphics[width=0.24\textwidth]{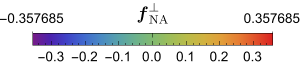}};
		\node(NFP1Bar)	[right=of NFM12Bar]		{\includegraphics[width=0.24\textwidth]{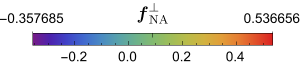}};
		\node(NFM1Bar)	[right=of NFP1Bar]		{\includegraphics[width=0.24\textwidth]{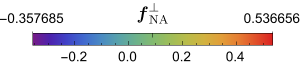}};
		\node(P12Note)	at ($(P12.center)!0.55!(P12.north west)$)	{$+\frac{1}{2}$:};
		\node(M12Note)	at ($(M12.center)!0.55!(M12.north west)$)	{$-\frac{1}{2}$:};
		\node(P1Note)	at ($(P1.center)!0.55!(P1.north west)$)		{$+1$:};
		\node(M1Note)	at ($(M1.center)!0.55!(M1.north west)$)		{$-1$:};
	\end{tikzpicture}
	\caption{Nematic fields, the resulting active nematic force fields and their normal part (top to bottom) for flat degenerated Q-tensor fields $ \qb\in\tangentQS $  
		bearing a $ +\frac{1}{2} $, $ -\frac{1}{2} $, $ +1 $ or $ -1 $-defect (left to right) at an elliptic point on a dome-shaped surface.
		The active nematic coefficient is given by $ \cNA = -\txi\cAC = 1 > 0 $ and results into a contractile force. 
		The Q-tensor field follows the quadratic ansatz $ \qb= S(\pb\otimes\pb-\frac{1}{2}\IdS) $ with a convenient and normalized director field $ \pb\in\tangentS $ 
		and scalar order field $ S=1-e^{-100 \hat{y}^2} $ (graph) where $ \hat{y}^2:=(y^1)^2 + (y^2)^2 $ \wrt\ the surface parametrization 
		$ \para(y^1,y^2) = [y^1, y^2, 1-(y^1)^2 - \frac{1}{2}(y^2)^2]^T $.
		The force field is orthogonally decomposed by $ \FbNA = \fb_{\NA} + f^{\bot}_{\NA}\normal $.
		The bottom row shows only the normal part $ f^{\bot}_{\NA}\normal $.}
	\label{fig:nematic_activity_hill}
\end{figure}

\begin{figure}[htp]
	\centering
	\begin{tikzpicture}[node distance=0pt]
		\node(P12) 	at (0,0) 		{\includegraphics[width=0.24\textwidth]{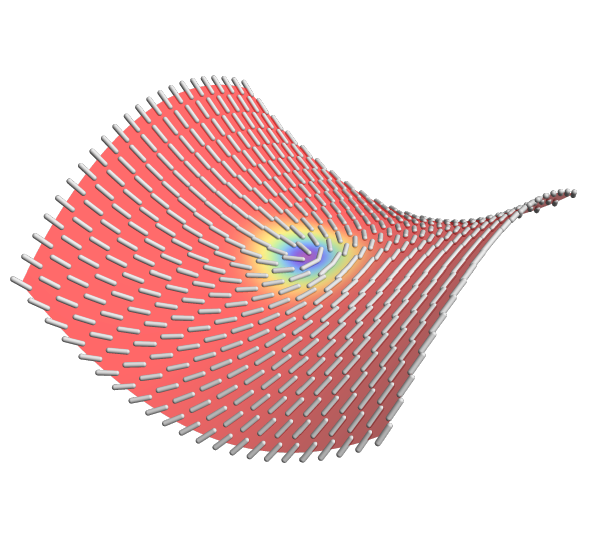}};
		\node(M12) 	[right=of P12] 	{\includegraphics[width=0.24\textwidth]{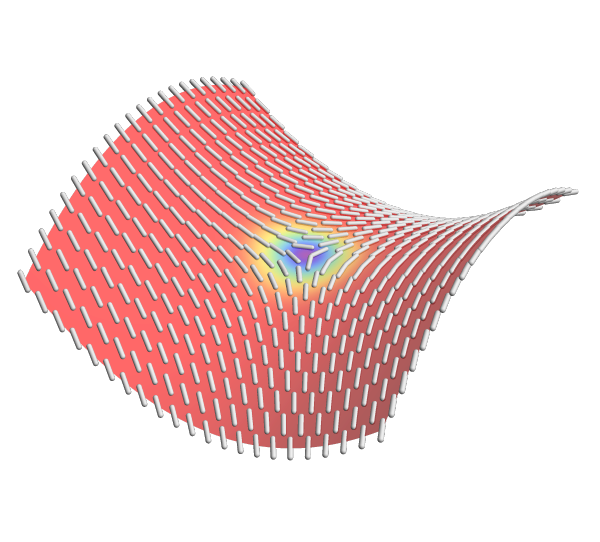}};
		\node(P1) 	[right=of M12]  {\includegraphics[width=0.24\textwidth]{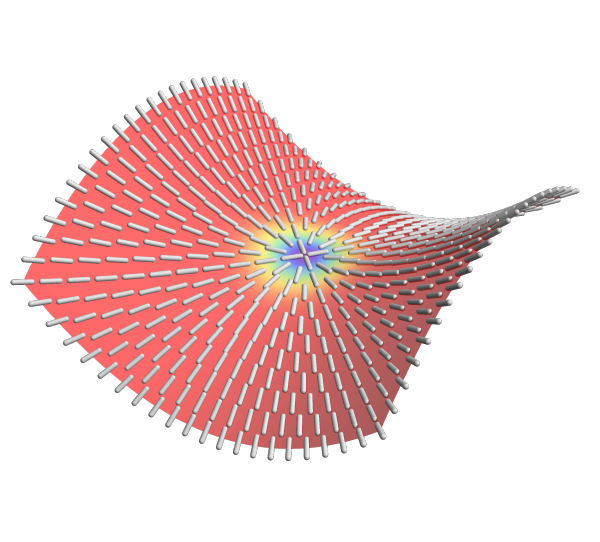}};
		\node(M1) 	[right=of P1]  	{\includegraphics[width=0.24\textwidth]{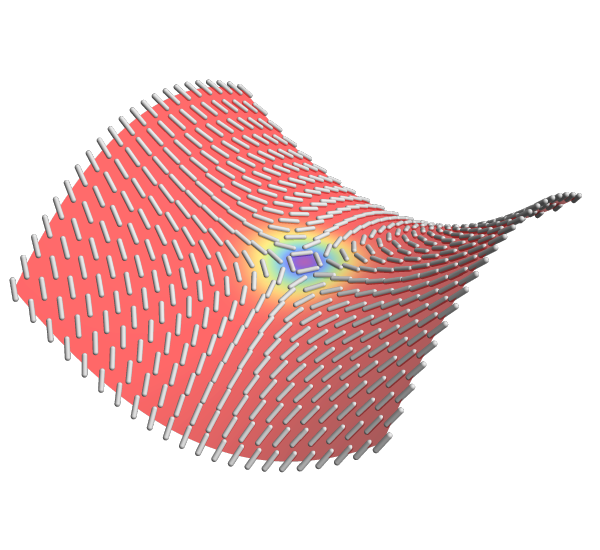}};
		\node(FP12)	[below=20pt of P12]	{\includegraphics[width=0.24\textwidth]{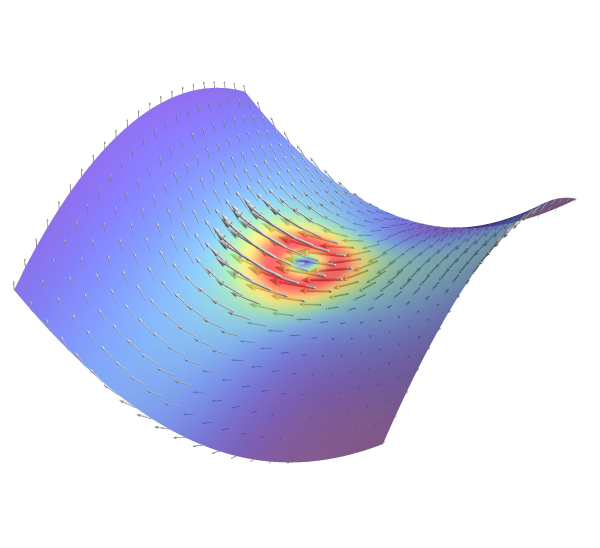}};
		\node(FM12)	[right=of FP12]		{\includegraphics[width=0.24\textwidth]{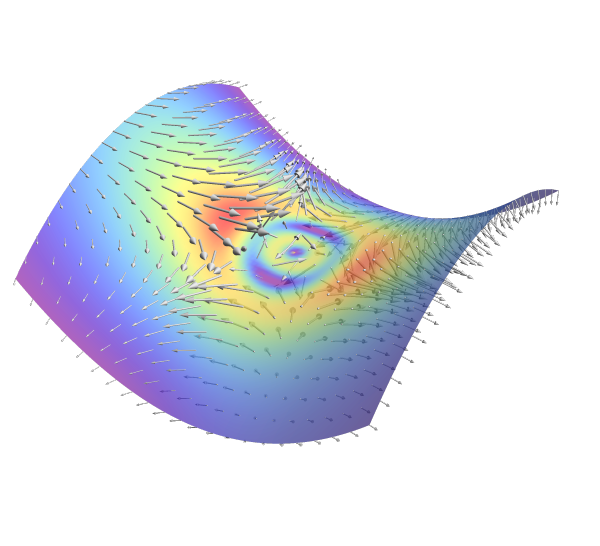}};
		\node(FP1) 	[right=of FM12]		{\includegraphics[width=0.24\textwidth]{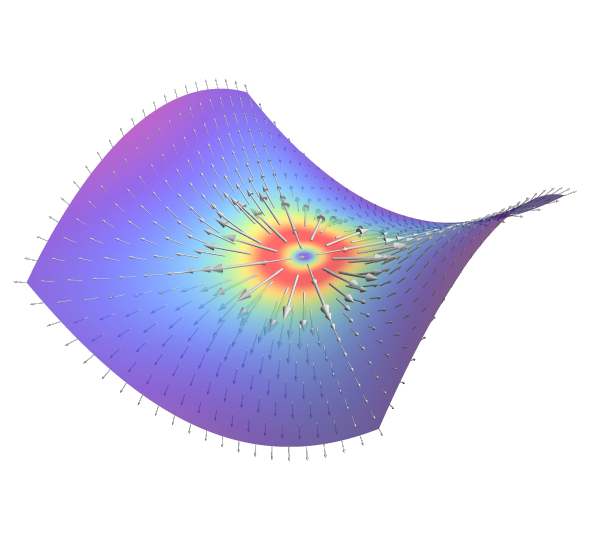}};
		\node(FM1) 	[right=of FP1]  	{\includegraphics[width=0.24\textwidth]{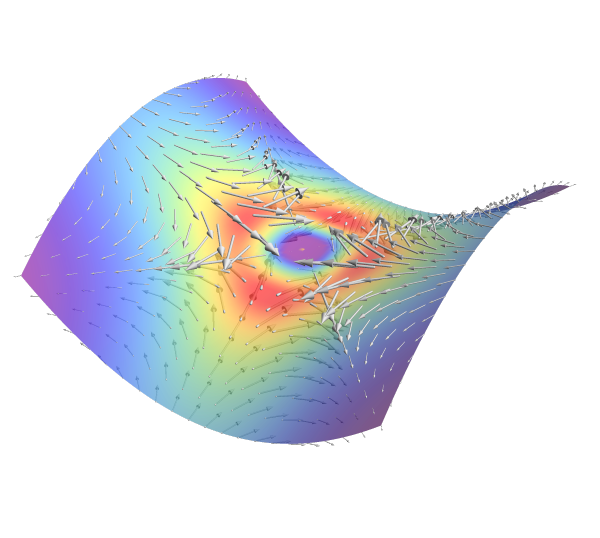}};
		\node(NFP12)	[below=of FP12]		{\includegraphics[width=0.24\textwidth]{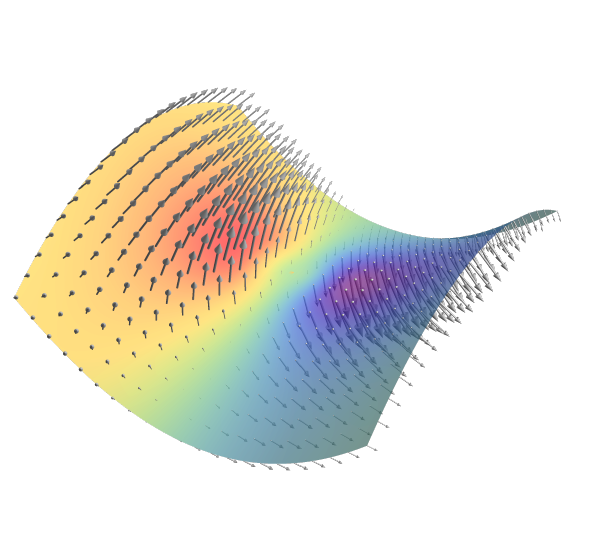}};
		\node(NFM12)	[right=of NFP12]	{\includegraphics[width=0.24\textwidth]{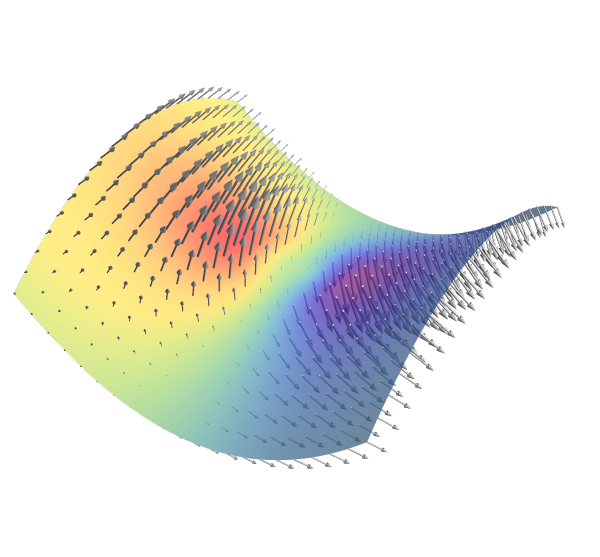}};
		\node(NFP1) 	[right=of NFM12]	{\includegraphics[width=0.24\textwidth]{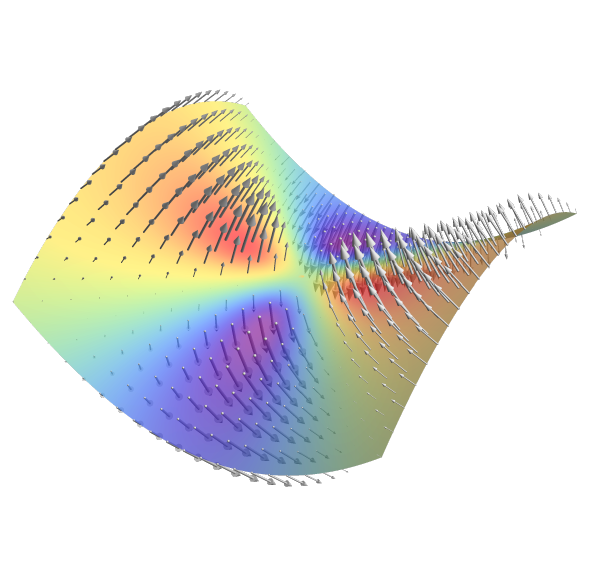}};
		\node(NFM1) 	[right=of NFP1]  	{\includegraphics[width=0.24\textwidth]{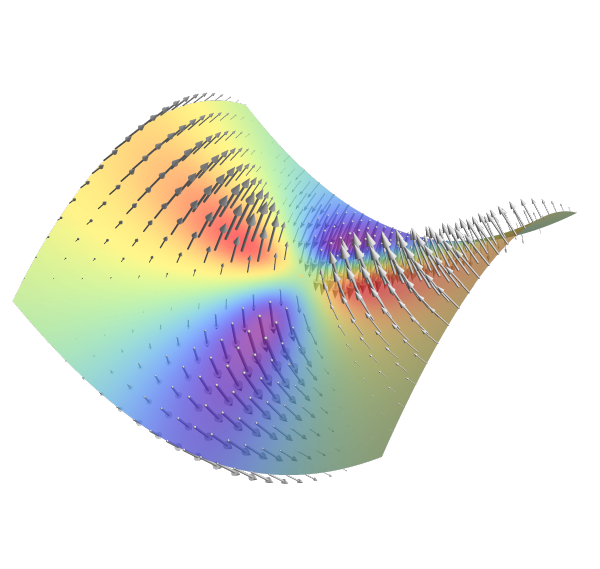}};
		\node(SBar) [below=-7pt of P12, xshift=0.12\textwidth] {\includegraphics[width=0.24\textwidth]{IMG/orderbarlegend.pdf}};
		\node(SBarGraph) [below=-7pt of P1, xshift=0.10\textwidth] {\includegraphics[width=0.5\textwidth]{IMG/order_distance_graph.pdf}};
		\node(FP12Bar)	[below=-14pt of FP12]	{\includegraphics[width=0.24\textwidth]{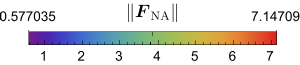}};
		\node(FM12Bar)	[right=of FP12Bar]		{\includegraphics[width=0.24\textwidth]{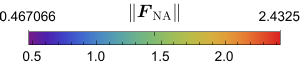}};
		\node(FP1Bar)	[right=of FM12Bar]		{\includegraphics[width=0.24\textwidth]{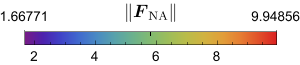}};
		\node(FM1Bar)	[right=of FP1Bar]		{\includegraphics[width=0.24\textwidth]{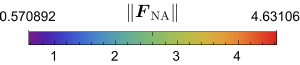}};
		\node(NFP12Bar)	[below=-14pt of NFP12]	{\includegraphics[width=0.24\textwidth]{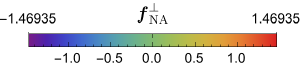}};
		\node(NFM12Bar)	[right=of NFP12Bar]		{\includegraphics[width=0.24\textwidth]{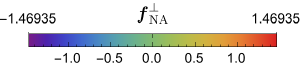}};
		\node(NFP1Bar)	[right=of NFM12Bar]		{\includegraphics[width=0.24\textwidth]{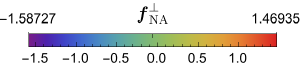}};
		\node(NFM1Bar)	[right=of NFP1Bar]		{\includegraphics[width=0.24\textwidth]{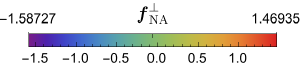}};
		\node(P12Note)	at ($(P12.center)!0.7!(P12.north west)$)	{$+\frac{1}{2}$:};
		\node(M12Note)	at ($(M12.center)!0.7!(M12.north west)$)	{$-\frac{1}{2}$:};
		\node(P1Note)	at ($(P1.center)!0.7!(P1.north west)$)		{$+1$:};
		\node(M1Note)	at ($(M1.center)!0.7!(M1.north west)$)		{$-1$:};
	\end{tikzpicture}
	\caption{Nematic fields, the resulting active nematic force fields and their normal part (top to bottom) for flat degenerated Q-tensor fields $ \qb\in\tangentQS $  
                bearing a $ +\frac{1}{2} $, $ -\frac{1}{2} $, $ +1 $ or $ -1 $-defect (left to right) at a saddle point.
                The active nematic coefficient is given by $ \cNA = -\txi\cAC = 1 > 0 $ and results into a contractile force. 
                The Q-tensor field follows the quadratic ansatz $ \qb= S(\pb\otimes\pb-\frac{1}{2}\IdS) $ with a convenient and normalized director field $ \pb\in\tangentS $ 
                and scalar order field $ S=1-e^{-100 \hat{y}^2} $ (graph) where $ \hat{y}^2:=(y^1)^2 + (y^2)^2 $ \wrt\ the surface parametrization 
                $ \para(y^1,y^2) = [y^1, y^2, 1+(y^1)^2 - \frac{3}{4}(y^2)^2]^T $.
                The force field is orthogonally decomposed by $ \FbNA = \fb_{\NA} + f^{\bot}_{\NA}\normal $.
                The bottom row shows only the normal part $ f^{\bot}_{\NA}\normal $.}
	\label{fig:nematic_activity_saddle}
\end{figure}

The considered special cases in Sections \ref{sec:existing} all have been treated numerically. We here refrain from numerically solving the derived active surface Beris-Edwards models, Eqs. \eqref{eq:model_lagrange_multiplier_a} or \eqref{eq:model_conforming}, to explore the full coupling between the surface Q-tensor field, the shape of the surface and the flow field. Instead we analyse the impact of the active nematic force at topological defects in the Q-tensor field. 

One of the celebrated properties of active nematodynamics in two dimensions is the characteristic flow fields at $+1/2$ and $-1/2$ defects. In \cite{Giomi_2014,Giomi_2015} it has been demonstrated that depending on the sign of the active stress $+1/2$ defects self-propel in the direction of their ‘tail’ (contractile) or ‘head’ (extensile), while $-1/2$ defects are only passively transported with the flow. This behaviour even allows to model active nematodynamics in a coarse-grained perspective as interacting self-propelled particles \cite{shankar2018defect,Doostmohammadi_2018,shankar2022topological}. While this analogy still holds if instead of a two dimensional setting a spherical surface is considered, see e.g. \cite{Keberetal_Science_2014}, it breaks down on geometrically stationary surfaces with varying mean curvature \cite{Alaimoetal_SR_2017,henkes2018dynamical,Nestler_2022}. This already indicates the strong influence of curvature effects at defects. Allowing for surface deformations changes the picture even more drastically. In Figures \ref{fig:nematic_activity_hill} and \ref{fig:nematic_activity_saddle} the nematic fields, the resulting active nematic force fields and their normal parts are shown for flat degenerate Q-tensor fields $ \qb\in\tangentQS $ in the vicinity of $ +1/2 $, $ -1/2 $, $ +1 $ and $ -1 $ defects on different geometries. 

Under the presence of curvature all defects become active and contribute not only tangential forces but also normal forces. Even if the magnitude of these normal forces is significantly lower, this confirms that the proposed role of topological defects in morphological changes can be modeled by the derived active surface Beris-Edwards models. 

Integer defects, $+ 1$ and $- 1$ defects, such as vortices, spirals or asters indeed have been found in nature at tissue scales. They colocalize with the
mouth, tentacles and foot of hydra during its development \cite{Maroudas-Sacksetal_NP_2021}. Using myoblasts it was also shown that integer topological defects can generate force gradients that concentrate compressive stresses and force cellular mounds to grow into cylindrical protrusions \cite{guillamat2022integer}. The active nematic force fields shown in Figures 
\ref{fig:nematic_activity_hill} and \ref{fig:nematic_activity_saddle} indeed differ between defect types. The once for $+ 1$ defects lead to similar structures as the once obtained in \cite{Vafaetal_PRL_2022,Al_Izzi_2023,pearce2023passive} and \cite{guillamat2022integer}, which further confirms the validity of the proposed model. However, the role of these forces in the dynamic evolution and the interplay of the forces generated by different defects as well as the other coupling mechanisms can only be explored by solving the full active surface Beris-Edwards equations Eqs. \eqref{eq:model_lagrange_multiplier_a} or \eqref{eq:model_conforming}. 

Extending the models by adding concentration fields of morphogens, growth factors and other signalling molecules and coupling the nematic field to their gradients, as in \cite{Wangetal_PNAS_2023} in a stationary setting, will lead to a dynamic theory of morphogenesis. To realize this and solve the equations, not only requires advanced computational approaches to deal with the highly non-linear vector- and tensor-valued surface partial differential equations, e.g. surface finite elements \cite{nestler2019finite} and higher order surface approximations \cite{praetorius2022dune}, but also supportive experimental studies to determine the mechanical properties of the tissue in order to reduce the complexity of the model by distinguishing significant and less significant coupling terms in the proposed equations. Thus before, e.g. the dynamics of gastrulation in Drosophila \cite{munster2019attachment}, can be computationally reproduced with the proposed approach, experiments which focus on specific aspects regarding the dynamics of geometric coupling or immobility mechanisms are required.

\appendix

\section{Quantities of the Passive Model} \label{app:a}

In the following subsections all quantities of the passive model are listed in tabular form.
The derivation and explanations of these terms can be found in \cite{Nitschke2023}.

\subsection{Necessary Quantities of the General Model}

\begin{table}[H]
    \centering
    \renewcommand{\arraystretch}{1.3}
    \begin{tabular}{|c|c|c|}
    \hline
        \multicolumn{2}{|c|}{Identifier}
                & Expression\\
    \hline
        \multicolumn{2}{|c|}{$ \Dmat\Vb $}
                & $ \partial_t\Vb + (\nablaC\Vb)(\Vb-\Vb_{\!\ofrak}) $ \\
    \hline
        \multirow{2}{*}{$ \Dphi\Qb $}
            & $ \Phi=\jau $
                & $ \partial_t\Qb + (\nablaC\Qb)(\Vb-\Vb_{\!\ofrak}) - \Abcal[\Vb]\Qb + \Qb\Abcal[\Vb] $\\
    \cline{2-3}
            & $ \Phi=\mfrak $
                & $ \partial_t\Qb + (\nablaC\Qb)(\Vb-\Vb_{\!\ofrak}) $ \\
    \hline
        \multicolumn{2}{|c|}{$ \SigmabEL $}
                & $ -L\left( (\nablaC\Qb)^{T} \dbdot\nablaC\Qb - \frac{\| \nablaC\Qb \|^2}{2}\IdS \right) $\\
    \hline
        \multicolumn{2}{|c|}{$ \HbEL $}
                & $ L \DeltaC\Qb $\\
    \hline
        \multicolumn{2}{|c|}{$ \pTH $}
                & $  a \Tr\Qb^2 + \frac{2b}{3}\Tr\Qb^3 + c\Tr\Qb^4 $ \\
    \hline
        \multicolumn{2}{|c|}{$ \HbTH $}
                & $ -2 \left( a \Qb + b\left( \Qb^2 - \frac{\Tr\Qb^2}{3}\Id \right) + c\Tr(\Qb^2)\Qb \right) $\\
    \hline
        \multicolumn{2}{|c|}{$ \fnorBE $}
                & $ -\kappa\left( \DeltaC\meanc + \left( \meanc - \meanc_{0} \right)\left( \frac{1}{2}\meanc(\meanc + \meanc_{0}) - 2\gaussc  \right) \right) $ \\
    \hline
        \multirow{2}{*}{$ \SigmabIM $}
            & $ \Phi=\jau $
                & $ M (\Id + \normal\otimes\normal)(\Qb\Djau\Qb - (\Djau\Qb)\Qb)\IdS $\\
    \cline{2-3}
            & $ \Phi=\mfrak $
                & $ \nullb $ \\
    \hline
        \multicolumn{2}{|c|}{$ \SigmabNV^0 $}
                & $ 2\coeffIF\Sb[\Vb] $\\
    \hline
        \multirow{2}{*}{$ \SigmabNV^1 $}
            &  $ (\Phi=\jau) $
                & $ -\coeffIF\left( \IdS(\Djau\Qb)\IdS + (3\IdS+2\normal\otimes\normal)\Qb\Sb[\Vb] + \Sb[\Vb]\Qb\IdS \right) $\\
    \cline{2-3}
            & $ (\Phi=\mfrak) $
                & $ -\coeffIF\left( \IdS(\Dmat\Qb)\IdS + \IdS\Qb\nablaC\Vb + 2\Qb\Sb[\Vb] + (\nablaC\Vb)^T\Qb\IdS \right)  $ \\
    \hline
        \multicolumn{2}{|c|}{$ \HbNV^1 $}
                & $ \coeffIF\Sb[\Vb] $ \\
    \hline
        \multirow{2}{*}{$ \SigmabNV^2 $}
            &  $ (\Phi=\jau) $
                & $ \coeffIF\left( \Qb\Djau\Qb\IdS - \normal\otimes \IdS \Qb (\Djau\Qb) \normal + \IdS\Qb\Sb[\Vb]\Qb\IdS + \Qb^2\Sb[\Vb] \right) $ \\
    \cline{2-3}
            & $ (\Phi=\mfrak) $
                & \splitcelltab{$ \coeffIF\big( \Qb\Dmat\Qb\IdS -\normal\otimes\IdS\Qb\left( (\Dmat\Qb)\normal - 2\Abcal[\Vb]\Qb\normal - \Qb(\nablaC\Vb)^T\normal \right) 
                                                +  \IdS\Qb\Gbcal^{T}[\Vb]\Qb\IdS + \Qb^2\nablaC\Vb \big)$} \\
    \hline
        \multirow{2}{*}{$ \widetilde{\Hb}^{2,\Phi}_{\NV} $}
            & $ \Phi=\jau $
                & $ -\frac{\coeffIF}{2} \left( \Qb\Sb[\Vb] + \Sb[\Vb]\Qb - \frac{\Qb\dbdot\Sb[\Vb]}{3}\Id \right) $ \\
    \cline{2-3}
            & $ \Phi=\mfrak $
                & $ -\frac{\coeffIF}{2} \left( \Qb\Gbcal[\Vb] + \Gbcal^T[\Vb]\Qb - \frac{\Qb\dbdot\Sb[\Vb]}{3}\Id \right) $ \\
    \hline
    \end{tabular}
    \caption{Necessary terms for \eqref{eq:model_lagrange_multiplier}, \resp\ \eqref{eq:model_lagrange_multiplier_a}, in the Surface Beris-Edwards models for a consistent choice $ \Phi\in\{\jau,\mfrak\} $.
            The deformation gradient field is given by $ \Gbcal[\Vb] = \nablaC\Wb - \normal\nablaC\Wb\otimes\normal $  \eqref{eq:Gbcal}. $ \Sb[\Vb] $ \eqref{eq:Sb} and $\Abcal[\Vb]$ \eqref{eq:Abcal} are its symmetric and skew-symmetric parts.
            Time derivatives are determined \wrt\ an observer velocity $ \Vb_{\!\ofrak}\in\tangentR $. 
            The choice of $ \Phi $ for $  \SigmabNV^1  $ and $  \SigmabNV^2  $ is optional. Both given representations state equal tensor fields.}
    \label{tab:forces}
\end{table}

\subsection{Optional Constrain Quantities of the General Model}

\begin{table}[H]
	\centering
	\renewcommand{\arraystretch}{1.2}
	\begin{tabular}{|c|c|c|c|c|}
		\hline
		$ \gamma $
		& $\Fb_{\gamma}$
		& $ \Hb_{\gamma} $
		& $  \Cb_{\gamma} $ & $ \Vcal_{\gamma} $.\\
		\hline
		$ \SC $ 
		& \splitcelltab{$ \DivC\big(\Qb(\normal,\normal) \normal \otimes  \lambdabSC $ \\ $-\normal \otimes \IdS\Qb\lambdabSC \big)$ } 
		& $ -\frac{1}{2}\left( \lambdabSC\otimes\normal + \normal\otimes\lambdabSC \right) $ 
		& $ \IdS\Qb\normal $ & $ \tangentS $\\
		\hline
		$ \CB $
		& $ \nullb $
		& $ -\lambdaCB \left(\normal\otimes\normal -\frac{1}{3}\Id \right)$
		& $\Qb(\normal,\normal) - \beta_0 $ & $ \tangentS[^0] $\\
		\hline
		$ \UN $
		& $\nullb$
		& \splitcelltab{$6\projQR\left( \LambdabUN\Qb^3 + \Qb\LambdabUN\Qb^2 \right)$\\
			$-5(\Tr\Qb^2) \projQR\left( \LambdabUN\Qb \right)
			-5 (\LambdabUN\dbdot\Qb^2)\Qb$}
		& \splitcelltab{$  \Qb^4 - \frac{5}{6}(\Tr\Qb^2)\Qb^2$\\$\quad+ \frac{1}{9}(\Tr\Qb^2)^2\Id $} & $ \tangentQR $\\
		\hline
		$ \IS $
		& $\nullb$
		& $\LambdabIS$
		& $ \Qb $ & $ \tangentQR $ \\
		\hline
		$ \NN $
		& $\lambdaNN \normal$
		& $\nullb$
		& $ \Vb\normal $ & $ \tangentS[^0] $ \\
		\hline
		$ \NF $
		& $ \LambdabNF $
		& $\nullb$
		& $ \Vb $ & $\tangentR$ \\
		\hline
	\end{tabular}
	\caption{Generalized constraint forces $ \Fb_{\gamma}\in\tangentR $, $  \Hb_{\gamma}\in\tangentQR $ and $  \Cb_{\gamma}\in\Vcal_{\gamma} $ \wrt\ Lagrange parameter in $\Vcal_{\gamma} $. 
		These terms apply in the Surface Beris-Edwards models \eqref{eq:model_lagrange_multiplier} in some circumstances.
	}
	\label{tab:constraints}
\end{table}

\subsection{Necessary Quantities of the Surface Conforming Model}

\begin{table}[H]
    \centering
    \renewcommand{\arraystretch}{1.3}
    \begin{tabular}{|c|c|c|}
    \hline
        \multicolumn{2}{|c|}{Identifier}
                & Expression \\
    \hline
        \multicolumn{2}{|c|}{$ \ab $}
                & $ \IdS\Dmat\Vb = (\partial_t v^i)\partial_i\para_{\!\ofrak} + \nabla_{\vb-\vb_{\!\ofrak}}\vb + \nabla_{\vb}\vb_{\!\ofrak} - \vnor\left( \nabla\vnor + 2\shop\vb \right) $\\
    \hline
        \multicolumn{2}{|c|}{$ \anor $}
                & $ \normal\Dmat\Vb = \partial_t\vnor + \nabla_{2\vb-\vb_{\!\ofrak}}\vnor + \shop(\vb,\vb) $ \\
    \hline
        \multirow{2}{*}{$ \timeD\qb $}
            & $ \Phi=\jau $
                & $ \timeJ\qb = \dot{\qb} - \Ab[\Vb]\qb + \qb\Ab[\Vb] $ \\
    \cline{2-3}
            & $ \Phi=\mfrak $
                & $ \dot{\qb} = (\partial_t q^{ij})\partial_i\para_{\!\ofrak}\otimes\partial_j\para_{\!\ofrak} + \nabla_{\vb-\vb_{\!\ofrak}}\qb + \Gb[\Vb_{\!\ofrak}]\qb + \qb\Gb^T[\Vb_{\!\ofrak}] $ \\
    \hline
        \multicolumn{2}{|c|}{$ \sigmabEL $}
                & \splitcelltab{$ -L\big( (\nabla\qb)^{T} \dbdot\nabla\qb + \frac{3}{2}\nabla\beta\otimes\nabla\beta 
                                          - \frac{1}{4}\left( 2\normsq{\tangentS[^3]}{\nabla\qb} + 3\normsq{\tangentS[^2]}{\nabla\beta}  \right)\IdS  $ \\
                                $ - 6\gaussc\beta\qb + \frac{1}{2}\left( 2\meanc\Tr\qb^2 - 12\beta \qb \dbdot \shop  + 9 \meanc\beta^2 \right)\left( \shop - \frac{\meanc}{2}\IdS \right)\big) $} \\
    \hline
        \multicolumn{2}{|c|}{$ \zetabEL $}
                & $ L\left( 2(\nabla\qb)\dbdot\shop + \qb\nabla\meanc - 3\shop\nabla\beta - \frac{3}{2}\beta\nabla\meanc \right) $ \\
    \hline
        \multicolumn{2}{|c|}{$ \hbEL $}
                & $ L \left(  \Delta\qb - (\meanc^2-2\gaussc)\qb + 3\beta \meanc \left( \shop - \frac{\meanc}{2}\IdS \right) \right) $\\
    \hline
        \multicolumn{2}{|c|}{$ \omegaEL $}
                & $ L \left( \Delta\beta + 2 \meanc\shop \dbdot \qb  - 3\beta \left( \meanc^2 - 2\gaussc \right)  \right) $ \\
    \hline
        \multicolumn{2}{|c|}{$ \pTH $}
                & $ \frac{1}{2}\left( 2a - 2b\beta + c\left( \Tr\qb^2 + 3\beta^2 \right) \right)\Tr\qb^2
                     +\frac{1}{8}\left( 12 a + 4b\beta + 9c\beta^2 \right)\beta^2 $\\
    \hline
        \multicolumn{2}{|c|}{$ \hbTH $}
                & $ -\left(2a - 2b\beta + 3c\beta^2  + 2c\Tr\qb^2\right)\qb $\\
    \hline
        \multicolumn{2}{|c|}{$ \omegaTH $}
                & $ -\left(2a + b\beta + 3c\beta^2  + 2c\Tr\qb^2\right)\beta + \frac{2}{3} b \Tr\qb^2 $ \\
    \hline
        \multicolumn{2}{|c|}{$ \fnorBE $}
                & $ -\kappa\left( \Delta\meanc + \left( \meanc - \meanc_{0} \right)\left( \frac{1}{2}\meanc(\meanc + \meanc_{0}) - 2\gaussc  \right) \right) $\\
    \hline
        \multirow{2}{*}{$ \sigmabIM $}
            & $ \Phi=\jau $
                & $ M \left( \qb\timeJ\qb - (\timeJ\qb)\qb \right) $\\
    \cline{2-3}
            & $ \Phi=\mfrak $
                & $ \nullb $ \\
    \hline
        \multirow{2}{*}{$ \zetabIM $}
            & $ \Phi=\jau $
                & $ \nullb $ \\
    \cline{2-3}
            & $ \Phi=\mfrak $
                & $  -M \left(\nabla\vnor + \vb\shop\right)\left( \qb - \frac{3}{2}\beta\IdS \right) $  \\
    \hline
        \multicolumn{2}{|c|}{$ \sigmabNV^0 $}
                & $ 2\coeffIF\Sb[\Vb] $\\
    \hline
        \multirow{2}{*}{$ \sigmabNV^1 $}
            &  $ (\Phi=\jau) $
                & $ -\coeffIF\left(\timeJ\qb - \frac{\dot{\beta}}{2}\IdS + 3\qb\Sb[\Vb] + \Sb[\Vb]\qb - 2\beta\Sb[\Vb] \right) $\\
    \cline{2-3}
            & $ (\Phi=\mfrak) $
                & $ -\coeffIF\left( \dot{\qb} - \frac{\dot{\beta}}{2}\IdS + \qb(2\Gb[\Vb]+\Gb^T[\Vb]) + \Gb^T[\Vb]\qb - 2\beta\Sb[\Vb] \right)  $ \\
    \hline
        \multicolumn{2}{|c|}{$ \hbNV^1 $}
                & $ \coeffIF\Sb[\Vb] $ \\
    \hline
        \multirow{2}{*}{$ \sigmabNV^2 $}
            &  $ (\Phi=\jau) $
                & \splitcelltab{$ \coeffIF\big( \qb\timeJ\qb - \frac{1}{2}\left( \beta\timeJ\qb + \dot{\beta}\qb \right)
                        + \frac{1}{4}\beta\dot{\beta}\IdS + \qb\Sb[\Vb]\qb$ \\
                  $-\frac{1}{2}\beta\left( 3\qb\Sb[\Vb] + \Sb[\Vb]\qb \right)  
                    + \frac{1}{2}\left( \Tr\qb^2 + \beta^2 \right)\Sb[\Vb] \big) $} \\
    \cline{2-3}
            & $ (\Phi=\mfrak) $
                & \splitcelltab{$ \coeffIF\big(  \qb\dot{\qb} - \frac{1}{2}\left( \beta\dot{\qb} + \dot{\beta}\qb \right) 
                          + \frac{1}{4}\beta\dot{\beta}\IdS + \qb\Gb^T[\Vb]\qb$ \\
                  $ -\frac{1}{2}\beta\left( \qb(2\Gb[\Vb]+\Gb^T[\Vb]) + \Gb^T[\Vb]\qb \right)  
                  + \frac{1}{2}(\Tr\qb^2)\Gb[\Vb] + \frac{1}{2}\beta^2\Sb[\Vb]\big) $} \\
    \hline
        \multirow{2}{*}{$ \widetilde{\hb}^{2,\Phi}_{\NV} $}
            & $ \Phi=\jau $
                & $ -\frac{\coeffIF}{2} \left( \qb\Sb[\Vb] + \Sb[\Vb]\qb - (\qb\dbdot\Gb[\Vb])\IdS - \beta\Sb[\Vb]\right) $ \\
    \cline{2-3}
            & $ \Phi=\mfrak $
                & $ -\frac{\coeffIF}{2} \left( \qb\Gb[\Vb] + \Gb^T[\Vb]\qb -(\qb\dbdot\Gb[\Vb])\IdS - \beta\Sb[\Vb] \right) $ \\
    \hline
        \multicolumn{2}{|c|}{$ \widetilde{\omega}^{2}_{\NV} $}
                & $ \frac{\coeffIF}{3} \qb\dbdot\Gb[\Vb] $ \\
    \hline
    \end{tabular}
    \caption{Necessary terms for the Surface Conforming Beris-Edwards models \eqref{eq:model_conforming} for a consistent choice $ \Phi\in\{\jau,\mfrak\} $.
            These representations comprise the tangential deformation gradient $ \Gb[\Vb] = \nabla\vb - \vnor\shop $ \eqref{eq:Gb} of the material velocity $ \Vb=\vb+\vnor\normal $.
            $ \Sb[\Vb] $ \eqref{eq:Sb} and $\Ab[\Vb]$ \eqref{eq:Ab} are its symmetric and skew-symmetric part.
            Time derivatives are determined \wrt\ an observer velocity $ \Vb_{\!\ofrak}= \vb_{\ofrak} + \vnor\normal $. }
    \label{tab:forces_conforming}
\end{table}

\subsection{Optional Constrain Quantities of the Surface Conforming Model}

\begin{table}[H]
    \centering
    \renewcommand{\arraystretch}{1.2}
    \begin{tabular}{|c|c|c|c|c|}
        \hline
        $ \gamma $
            & $\left(\fb_{\gamma}, \fnor[\gamma]\right)$
                & $ \left(\hb_{\gamma}, \omega_{\gamma}\right) $
                    & $  \Cb_{\gamma} $ & $ \Vcal_{\gamma} $\\
        \hline
        $ \CB $
            & $ \left(\nullb, 0\right) $
                & $ \left(\nullb, -\frac{2}{3}\lambdaCB\right) $
                    & $\beta - \beta_0 $ & $ \tangentS[^0] $\\
        \hline
        $ \UN $
            & $\left(\nullb, 0\right)$
                & \splitcelltab{$\Big(-6\beta\qb\lambdabUN\qb + \frac{4}{3}(\Tr\qb^2)\projQS(\lambdabUN\qb)$\\
                          $ + \left( 5\beta\lambdabUN\dbdot\qb + \lambdabotUN\Tr\qb^2\right)\qb$\\
                          $ - \frac{1}{4}\beta\left( 14\Tr\qb^2 - 9\beta^2 \right)\lambdabUN$,\\
                          $ \frac{1}{3}(\Tr\qb^2)\left( 2\lambdabUN\dbdot\qb - 9\lambdabotUN\beta \right)\Big) $}
                    & \splitcelltab{$  \Big( \left( 2\Tr\qb^2 - 9\beta^2 \right)\beta\qb $,\\
                                $\left( 2\Tr\qb^2 - 9\beta^2 \right)\Tr\qb^2 \Big)$} & $ \tangentQS\times\tangentS[^0] $\\
        \hline
        $ \NN $
            & $\left(\nullb, \lambdaNN\right)$
                & $\left(\nullb, 0\right)$
                    & $ \vnor $ & $ \tangentS[^0] $ \\
        \hline
        $ \NF $
            & $ \left( \lambdab_{\NF}, \lambda^{\bot}_{\NF} \right) $
                & $\left(\nullb, 0\right)$
                    & $ \left( \vb, \vnor \right) $ & $\tangentS\times\tangentS[^0]$ \\
        \hline
    \end{tabular}
    \caption{Generalized constraint forces $ (\fb_{\gamma}, \fnor[\gamma])\in\tangentS\times\tangentS[^0]\cong\tangentR $, 
        $  (\hb_{\gamma}, \omega_{\gamma})\in\tangentQS\times\tangentS[^0] \cong \tangentCQR $ and $  \Cb_{\gamma}\in\Vcal_{\gamma} $ 
        \wrt\ Lagrange parameter in $ \Vcal_{\gamma} $. 
        These terms apply in the Surface Conforming Beris-Edwards model \eqref{eq:model_conforming} in some circumstances. For the constraints $ \UN $ and $ \NF $ we represent the Lagrange parameter by a pair of Lagrange parameters, \ie\ $ \LambdabUN\cong\left( \lambdabUN, \lambdabotUN \right) $
        and $ \LambdabNF\cong\left( \lambdab_{\NF}, \lambda^{\bot}_{\NF} \right) $, where $ \lambdabUN\in\tangentQS $, $ \lambdab_{\NF}\in\tangentS $ and $ \lambdabotUN,\lambda^{\bot}_{\NF}\in\tangentS[^0] $.
        }
    \label{tab:constraints_conforming}
\end{table}

\section{Flow Alignment}\label{sec:flow_alignment}

There exists different approaches in literature to align the nematic field with the flow. 
We aim to provide a very brief comparative overview in Table \ref{tab:flow_alignment} of the three most common approaches we encountered during our literature review.
The anisotropic part of the nematic viscosity reflects the anisotropic behavior of internal spatial distortion according to the apolar direction of the nematic field,
which yields flow alignment eventually.
Note that nematic viscosity results in a pure dissipative mechanism.
Another way to accomplish flow alignment is to involve a linear reaction term in compliance to the Onsager relations.
A third idea is to account the nematic field \wrt\ the extensional and rotational components of velocity gradients,
which can be seen as an improper extension of the Jaumann/corotational derivative which drives the immobility in context of \cite{Nitschke2023}.
Due to their different origins, they are only vaguely similar.
However, we believe that in many settings, especially near equilibrium, they yield similar solutions. 

\begin{table}[ht]
	\centering
	\renewcommand{\arraystretch}{1.2}
	\begin{tabular}{|c|c|c|c|}
	\hline
		Approach:	
			& Nematic Viscosity
				& Linear Reaction
					& Extended Corotation\\
	\hline
		$ \sigmabfa $
			& \splitcelltab{$ -\coeffIF\xi(\timeLl\qb + 2\qb\Sb[\vb]) + \coeffIF\xi^2\qb\timeLl\qb$\\
								$ = -\coeffIF\xi ( \frac{1}{M}\hb_\potenergy + \hat{\sigmab} ) + \landau(\xi^2)$}
				& $ \coeffIF_1 \hb_\potenergy $
					&\splitcelltab{$ -\txi \big( \hb_\potenergy + \hb_\potenergy\qb + \qb\hb_\potenergy $\\
									$- (\hb_\potenergy:\qb)(\IdS + 2\qb)\big)$} \\
	\hline
		$ \hbfa $
			& $ \frac{\coeffIF\xi}{M}\Sb[\vb] - \frac{\coeffIF\xi^2}{2M}\timeLQl\qb  $
				& $ -\coeffIF_1 \Sb[\vb] $
					& \splitcelltab{$ \txi \big( \Sb[\vb] + \Sb[\vb]\qb + \qb\Sb[\vb] $\\
									$- (\Sb[\vb]:\qb)(\IdS + 2\qb)\big)$}\\
	\hline
		See \eg:
			& \cite{Nitschke2023,Leslie_1966,Beris_1994,Parodi_1970}
				& \cite{Juelicher_2018,Kruse_2005,Salbreux_2009}
					& \cite{Rorai_2021,Thampi_2016,de_Gennes_book,Thijssen_2020}\\
	\hline
	\end{tabular}
	\caption{Juxtaposition of different flow alignment mechanism manifested in the fluid stress $  \sigmabfa\in\tangentS[^2] $ and molecular force $ \hbfa\in\tangentQS $
			for a flat-degenerated Q-tensor field $ \qb\in\tangentQS $ on a flat geometrical stationary surface.
			$ \xi,\coeffIF_1, \txi $ are the anisotropy/reactive coefficient/flow alignment parameter, though with different physical dimensions.
			The molecular field is given by $ \hb_\potenergy = L\Delta\qb -2 \left( a + c\Tr\qb^2 \right)\qb $, 
			$ \hat{\sigmab} = 2( \Sb[\vb]\qb + \qb\Sb[\vb]) $, 
			$ 2\Sb[\vb] = \timeLl\IdS = \nabla\vb + (\nabla\vb)^T $,
			and the lower-convected tangential Q-tensor rate of $ \qb $ by $ \timeLQl\qb = \timeLl\qb - (\Sb[\vb]:\qb)\IdS \in\tangentQS $, 
			where $ \timeLl\qb = \timeJ\qb + \Sb[\vb]\qb + \qb\Sb[\vb] $ is the lower-convected tangential rate of $ \qb $, see \cite{NitschkeVoigt_JoGaP_2022,NitschkeVoigt_2023}.}
	\label{tab:flow_alignment}
\end{table}

\noindent
{\textbf{Funding:}} A.V. was supported by DFG through FOR3013. \\
 
\noindent
{\textbf{Competing interests:}} There are no competing interests. \\
 
\noindent
{\textbf{Authors' contributions:}} This project was conceived by I.N. and A.V.; I.N. derived the theory, the results were analysed by I.N. and A.V. and the main text was written by I.N. and A.V..

\bibliographystyle{elsarticle-num}
\bibliography{NAC.bib}

\begin{thebibliography}{10}
\expandafter\ifx\csname url\endcsname\relax
  \def\url#1{\texttt{#1}}\fi
\expandafter\ifx\csname urlprefix\endcsname\relax\def\urlprefix{URL }\fi
\expandafter\ifx\csname href\endcsname\relax
  \def\href#1#2{#2} \def\path#1{#1}\fi

\bibitem{Maroudas-Sacksetal_NP_2021}
Y.~Maroudas-Sacks, L.~Garion, L.~Shani-Zerbib, A.~Livshits, E.~Braun, K.~Keren,
  Topological defects in the nematic order of actin fibres as organization
  centres of hydra morphogenesis, Nature Physics 17~(2) (2021) 251--259.
\newblock \href {https://doi.org/10.1038/s41567-020-01083-1}
  {\path{doi:10.1038/s41567-020-01083-1}}.

\bibitem{Vafaetal_PRL_2022}
F.~Vafa, L.~Mahadevan, Active nematic defects and epithelial morphogenesis,
  Physical Review Letters 129~(9) (2022) 098102.
\newblock \href {https://doi.org/10.1103/physrevlett.129.098102}
  {\path{doi:10.1103/physrevlett.129.098102}}.

\bibitem{Wangetal_PNAS_2023}
Z.~Wang, M.~C. Marchetti, F.~Brauns, Patterning of morphogenetic anisotropy
  fields, Proceedings of the National Academy of Sciences 120~(13) (2023)
  e2220167120.
\newblock \href {https://doi.org/10.1073/pnas.2220167120}
  {\path{doi:10.1073/pnas.2220167120}}.

\bibitem{saw2017topological}
T.~B. Saw, A.~Doostmohammadi, V.~Nier, L.~Kocgozlu, S.~Thampi, Y.~Toyama,
  P.~Marcq, C.~T. Lim, J.~M. Yeomans, B.~Ladoux, Topological defects in
  epithelia govern cell death and extrusion, Nature 544~(7649) (2017) 212--216.
\newblock \href {https://doi.org/10.1038/nature21718}
  {\path{doi:10.1038/nature21718}}.

\bibitem{de_Gennes_book}
P.~G. de~Gennes, J.~Prost, The Physics of Liquid Crystals, Second Edition,
  Clarendon Press, Oxford, 1993.
\newblock \href {https://doi.org/10.1093/oso/9780198520245.001.0001}
  {\path{doi:10.1093/oso/9780198520245.001.0001}}.

\bibitem{napoli2012surface}
G.~Napoli, L.~Vergori, Surface free energies for nematic shells, Physical
  Review E 85~(6) (2012) 061701.
\newblock \href {https://doi.org/10.1103/physreve.85.061701}
  {\path{doi:10.1103/physreve.85.061701}}.

\bibitem{golovaty2017dimension}
D.~Golovaty, J.~A. Montero, P.~Sternberg, Dimension reduction for the landau-de
  gennes model on curved nematic thin films, Journal of Nonlinear Science
  27~(6) (2017) 1905--1932.
\newblock \href {https://doi.org/10.1007/s00332-017-9390-5}
  {\path{doi:10.1007/s00332-017-9390-5}}.

\bibitem{Nitschke_2018}
I.~Nitschke, M.~Nestler, S.~Praetorius, H.~Löwen, A.~Voigt, Nematic liquid
  crystals on curved surfaces: a thin film limit, Proceedings of the Royal
  Society A: Mathematical, Physical and Engineering Sciences 474~(2214) (2018)
  20170686.
\newblock \href {https://doi.org/10.1098/rspa.2017.0686}
  {\path{doi:10.1098/rspa.2017.0686}}.

\bibitem{Park_EPL_1992}
J.~Park, T.~C. Lubensky, F.~C. MacKintosh, n-atic order and continuous shape
  changes of deformable surfaces of genus zero, Europhysics Letters ({EPL})
  20~(3) (1992) 279--284.
\newblock \href {https://doi.org/10.1209/0295-5075/20/3/015}
  {\path{doi:10.1209/0295-5075/20/3/015}}.

\bibitem{Nitschke_2020}
I.~Nitschke, S.~Reuther, A.~Voigt, Liquid crystals on deformable surfaces,
  Proceedings of the Royal Society A: Mathematical, Physical and Engineering
  Sciences 476~(2241) (2020).
\newblock \href {https://doi.org/10.1098/rspa.2020.0313}
  {\path{doi:10.1098/rspa.2020.0313}}.

\bibitem{Senoussi_2019}
A.~Senoussi, S.~Kashida, R.~Voituriez, J.-C. Galas, A.~Maitra,
  A.~Estevez-Torres, Tunable corrugated patterns in an active nematic sheet,
  Proceedings of the National Academy of Sciences 116~(45) (2019) 22464--22470.
\newblock \href {https://doi.org/10.1073/pnas.1912223116}
  {\path{doi:10.1073/pnas.1912223116}}.

\bibitem{Str_bing_2020}
T.~Strübing, A.~Khosravanizadeh, A.~Vilfan, E.~Bodenschatz, R.~Golestanian,
  I.~Guido, Wrinkling instability in 3d active nematics, Nano Letters 20~(9)
  (2020) 6281--6288.
\newblock \href {https://doi.org/10.1021/acs.nanolett.0c01546}
  {\path{doi:10.1021/acs.nanolett.0c01546}}.

\bibitem{Lehtinen2013}
O.~Lehtinen, S.~Kurasch, A.~Krasheninnikov, U.~Kaiser, Atomic scale study of
  the life cycle of a dislocation in graphene from birth to annihilation,
  Nature Communications 4~(1) (2013) 2098.
\newblock \href {https://doi.org/10.1038/ncomms3098}
  {\path{doi:10.1038/ncomms3098}}.

\bibitem{Zhang2014a}
T.~Zhang, X.~Li, H.~Gao, Defects controlled wrinkling and topological design in
  graphene, Journal of the Mechanics and Physics of Solids 67 (2014) 2--13.
\newblock \href {https://doi.org/10.1016/j.jmps.2014.02.005}
  {\path{doi:10.1016/j.jmps.2014.02.005}}.

\bibitem{benoitmaréchal2024mesoscale}
L.~Benoit-Maréchal, I.~Nitschke, A.~Voigt, M.~Salvalaglio, Mesoscale modeling
  of deformations and defects in crystalline sheets, arXiv (2024).
\newblock \href {https://doi.org/10.48550/arxiv.2309.11371}
  {\path{doi:10.48550/arxiv.2309.11371}}.

\bibitem{kawaguchi2017topological}
K.~Kawaguchi, R.~Kageyama, M.~Sano, Topological defects control collective
  dynamics in neural progenitor cell cultures, Nature 545~(7654) (2017)
  327--331.
\newblock \href {https://doi.org/10.1038/nature22321}
  {\path{doi:10.1038/nature22321}}.

\bibitem{guillamat2022integer}
P.~Guillamat, C.~Blanch-Mercader, G.~Pernollet, K.~Kruse, A.~Roux, Integer
  topological defects organize stresses driving tissue morphogenesis, Nature
  Materials 21~(5) (2022) 588--597.
\newblock \href {https://doi.org/10.1038/s41563-022-01194-5}
  {\path{doi:10.1038/s41563-022-01194-5}}.

\bibitem{Metselaar_2019}
L.~Metselaar, J.~M. Yeomans, A.~Doostmohammadi, Topology and morphology of
  self-deforming active shells, Physical Review Letters 123~(20) (Nov. 2019).
\newblock \href {https://doi.org/10.1103/physrevlett.123.208001}
  {\path{doi:10.1103/physrevlett.123.208001}}.

\bibitem{hoffmann2022theory}
L.~A. Hoffmann, L.~N. Carenza, J.~Eckert, L.~Giomi, Theory of defect-mediated
  morphogenesis, Science Advances 8~(15) (2022) eabk2712.
\newblock \href {https://doi.org/10.1126/sciadv.abk2712}
  {\path{doi:10.1126/sciadv.abk2712}}.

\bibitem{lubensky1992orientational}
T.~C. Lubensky, J.~Prost, Orientational order and vesicle shape, Journal de
  Physique II 2~(3) (1992) 371--382.
\newblock \href {https://doi.org/10.1051/jp2:1992133}
  {\path{doi:10.1051/jp2:1992133}}.

\bibitem{nelson2002toward}
D.~R. Nelson, Toward a tetravalent chemistry of colloids, Nano Letters 2~(10)
  (2002) 1125--1129.
\newblock \href {https://doi.org/10.1021/nl0202096}
  {\path{doi:10.1021/nl0202096}}.

\bibitem{Keberetal_Science_2014}
F.~C. Keber, E.~Loiseau, T.~Sanchez, S.~J. DeCamp, L.~Giomi, M.~J. Bowick,
  M.~C. Marchetti, Z.~Dogic, A.~R. Bausch, Topology and dynamics of active
  nematic vesicles, Science 345~(6201) (2014) 1135--1139.
\newblock \href {https://doi.org/10.1126/science.1254784}
  {\path{doi:10.1126/science.1254784}}.

\bibitem{Alaimoetal_SR_2017}
F.~Alaimo, C.~Köhler, A.~Voigt, Curvature controlled defect dynamics in
  topological active nematics, Scientific Reports 7~(1) (2017) 5211.
\newblock \href {https://doi.org/10.1038/s41598-017-05612-6}
  {\path{doi:10.1038/s41598-017-05612-6}}.

\bibitem{henkes2018dynamical}
S.~Henkes, M.~C. Marchetti, R.~Sknepnek, Dynamical patterns in nematic active
  matter on a sphere, Physical Review E 97~(4) (2018) 042605.
\newblock \href {https://doi.org/10.1103/physreve.97.042605}
  {\path{doi:10.1103/physreve.97.042605}}.

\bibitem{ellis2018curvature}
P.~W. Ellis, D.~J.~G. Pearce, Y.-W. Chang, G.~Goldsztein, L.~Giomi,
  A.~Fernandez-Nieves, Curvature-induced defect unbinding and dynamics in
  active nematic toroids, Nature Physics 14~(1) (2018) 85--90.
\newblock \href {https://doi.org/10.1038/nphys4276}
  {\path{doi:10.1038/nphys4276}}.

\bibitem{Nestler_2022}
M.~Nestler, A.~Voigt, Active nematodynamics on curved surfaces – the
  influence of geometric forces on motion patterns of topological defects,
  Communications in Computational Physics 31~(3) (2022) 947--965.
\newblock \href {https://doi.org/10.4208/cicp.oa-2021-0206}
  {\path{doi:10.4208/cicp.oa-2021-0206}}.

\bibitem{tanner2012coherent}
K.~Tanner, H.~Mori, R.~Mroue, A.~Bruni-Cardoso, M.~J. Bissell, Coherent angular
  motion in the establishment of multicellular architecture of glandular
  tissues, Proceedings of the National Academy of Sciences 109~(6) (2012)
  1973--1978.
\newblock \href {https://doi.org/10.1073/pnas.1119578109}
  {\path{doi:10.1073/pnas.1119578109}}.

\bibitem{wang2013rotational}
H.~Wang, S.~Lacoche, L.~Huang, B.~Xue, S.~K. Muthuswamy, Rotational motion
  during three-dimensional morphogenesis of mammary epithelial acini relates to
  laminin matrix assembly, Proceedings of the National Academy of Sciences
  110~(1) (2012) 163--168.
\newblock \href {https://doi.org/10.1073/pnas.1201141110}
  {\path{doi:10.1073/pnas.1201141110}}.

\bibitem{happel2022effects}
L.~Happel, D.~Wenzel, A.~Voigt, Effects of curvature on epithelial tissue
  —coordinated rotational movement and other spatiotemporal arrangements,
  Europhysics Letters 138~(6) (2022) 67002.
\newblock \href {https://doi.org/10.1209/0295-5075/ac757a}
  {\path{doi:10.1209/0295-5075/ac757a}}.

\bibitem{brandstatter2021curvature}
T.~Brandstätter, D.~B. Brückner, Y.~L. Han, R.~Alert, M.~Guo, C.~P.
  Broedersz, Curvature induces active velocity waves in rotating multicellular
  spheroids, arXiv (2021).
\newblock \href {https://doi.org/10.48550/arXiv.2110.14614}
  {\path{doi:10.48550/arXiv.2110.14614}}.

\bibitem{glentis2022emergence}
A.~Glentis, C.~Blanch-Mercader, L.~Balasubramaniam, T.~B. Saw,
  J.~d’Alessandro, S.~Janel, A.~Douanier, B.~Delaval, F.~Lafont, C.~T. Lim,
  D.~Delacour, J.~Prost, W.~Xi, B.~Ladoux, The emergence of spontaneous
  coordinated epithelial rotation on cylindrical curved surfaces, Science
  Advances 8~(37) (2022) eabn5406.
\newblock \href {https://doi.org/10.1126/sciadv.abn5406}
  {\path{doi:10.1126/sciadv.abn5406}}.

\bibitem{happel2024coordinated}
L.~Happel, A.~Voigt, Coordinated motion of epithelial layers on curved
  surfaces, Physical Review Letters 132~(7) (2024) 078401.
\newblock \href {https://doi.org/10.1103/physrevlett.132.078401}
  {\path{doi:10.1103/physrevlett.132.078401}}.

\bibitem{NitschkeSadikVoigt_A_2022}
I.~Nitschke, S.~Sadik, A.~Voigt, Tangential tensor fields on deformable
  surfaces -- {H}ow to derive consistent {$L^2$}-gradient flows, IMA Journal of
  Applied Mathematics 88~(6) (2023) 917--958.
\newblock \href {https://doi.org/10.1093/imamat/hxae006}
  {\path{doi:10.1093/imamat/hxae006}}.

\bibitem{stone2023note}
H.~A. Stone, M.~J. Shelley, E.~Boyko, A note about convected time derivatives
  for flows of complex fluids, Soft Matter 19~(28) (2023) 5353--5359.
\newblock \href {https://doi.org/10.1039/d3sm00497j}
  {\path{doi:10.1039/d3sm00497j}}.

\bibitem{NitschkeVoigt_2023}
I.~Nitschke, A.~Voigt, Tensorial time derivatives on moving surfaces: General
  concepts and a specific application for surface landau-de gennes models,
  Journal of Geometry and Physics 194 (2023) 105002.
\newblock \href {https://doi.org/10.1016/j.geomphys.2023.105002}
  {\path{doi:10.1016/j.geomphys.2023.105002}}.

\bibitem{Torres-Sanchez_2019}
A.~Torres-Sánchez, D.~Millán, M.~Arroyo, Modelling fluid deformable surfaces
  with an emphasis on biological interfaces, Journal of Fluid Mechanics 872
  (2019) 218--271.
\newblock \href {https://doi.org/10.1017/jfm.2019.341}
  {\path{doi:10.1017/jfm.2019.341}}.

\bibitem{Reuther_2020}
S.~Reuther, I.~Nitschke, A.~Voigt, A numerical approach for fluid deformable
  surfaces, Journal of Fluid Mechanics 900 (2020).
\newblock \href {https://doi.org/10.1017/jfm.2020.564}
  {\path{doi:10.1017/jfm.2020.564}}.

\bibitem{Krause_2023}
V.~Krause, A.~Voigt, A numerical approach for fluid deformable surfaces with
  conserved enclosed volume, Journal of Computational Physics 486 (2023)
  112097.
\newblock \href {https://doi.org/10.1016/j.jcp.2023.112097}
  {\path{doi:10.1016/j.jcp.2023.112097}}.

\bibitem{Helfrich}
W.~Helfrich, Elastic properties of lipid bilayers: Theory and possible
  experiments, Zeitschrift für Naturforschung C 28~(11–12) (1973) 693--703.
\newblock \href {https://doi.org/10.1515/znc-1973-11-1209}
  {\path{doi:10.1515/znc-1973-11-1209}}.

\bibitem{reymann2016cortical}
A.-C. Reymann, F.~Staniscia, A.~Erzberger, G.~Salbreux, S.~W. Grill, Cortical
  flow aligns actin filaments to form a furrow, eLife 5 (2016) e17807.
\newblock \href {https://doi.org/10.7554/eLife.17807}
  {\path{doi:10.7554/eLife.17807}}.

\bibitem{bhatnagar2023axis}
A.~Bhatnagar, M.~Nestler, P.~Gross, M.~Kramar, M.~Leaver, A.~Voigt, S.~W.
  Grill, Axis convergence in c. elegans embryos, Current Biology 33~(23)
  (2023) 5096--5108.
\newblock \href {https://doi.org/10.1016/j.cub.2023.10.050}
  {\path{doi:10.1016/j.cub.2023.10.050}}.

\bibitem{al2021active}
S.~C. Al-Izzi, R.~G. Morris, Active flows and deformable surfaces in
  development, in: Seminars in Cell \& Developmental Biology, Vol. 120,
  Elsevier, 2021, pp. 44--52.

\bibitem{salbreux2017mechanics}
G.~Salbreux, F.~Jülicher, Mechanics of active surfaces, Physical Review E
  96~(3) (2017) 032404.
\newblock \href {https://doi.org/10.1103/physreve.96.032404}
  {\path{doi:10.1103/physreve.96.032404}}.

\bibitem{Salbreux_2022}
G.~Salbreux, F.~Jülicher, J.~Prost, A.~Callan-Jones, Theory of nematic and
  polar active fluid surfaces, Physical Review Research 4~(3) (Aug. 2022).
\newblock \href {https://doi.org/10.1103/physrevresearch.4.033158}
  {\path{doi:10.1103/physrevresearch.4.033158}}.

\bibitem{Al_Izzi_2023}
S.~C. Al-Izzi, R.~G. Morris, Morphodynamics of active nematic fluid surfaces,
  Journal of Fluid Mechanics 957 (Feb. 2023).
\newblock \href {https://doi.org/10.1017/jfm.2023.18}
  {\path{doi:10.1017/jfm.2023.18}}.

\bibitem{Nitschke2023}
I.~Nitschke, A.~Voigt, {B}eris-{E}dwards models on evolving surfaces: A
  {L}agrange-{D}'{A}lembert approach, arXiv (2023).
\newblock \href {https://doi.org/10.48550/ARXIV.2311.06240}
  {\path{doi:10.48550/ARXIV.2311.06240}}.

\bibitem{BachiniKrauseNitschkeVoigt_2023}
E.~Bachini, V.~Krause, I.~Nitschke, A.~Voigt, Derivation and simulation of a
  two-phase fluid deformable surface model, Journal of Fluid Mechanics 977
  (2023).
\newblock \href {https://doi.org/10.1017/jfm.2023.943}
  {\path{doi:10.1017/jfm.2023.943}}.

\bibitem{NitschkeVoigt_JoGaP_2022}
I.~Nitschke, A.~Voigt, Observer-invariant time derivatives on moving surfaces,
  Journal of Geometry and Physics 173 (2022) 104428.
\newblock \href {https://doi.org/10.1016/j.geomphys.2021.104428}
  {\path{doi:10.1016/j.geomphys.2021.104428}}.

\bibitem{Arroyo_2009}
M.~Arroyo, A.~DeSimone, Relaxation dynamics of fluid membranes, Physical Review
  E 79~(3) (2009).
\newblock \href {https://doi.org/10.1103/physreve.79.031915}
  {\path{doi:10.1103/physreve.79.031915}}.

\bibitem{Reuther_2015}
S.~Reuther, A.~Voigt, The interplay of curvature and vortices in flow on curved
  surfaces, Multiscale Modeling {\&} Simulation 13~(2) (2015) 632--643.
\newblock \href {https://doi.org/10.1137/140971798}
  {\path{doi:10.1137/140971798}}.

\bibitem{Koba_2017}
H.~Koba, C.~Liu, Y.~Giga, Energetic variational approaches for incompressible
  fluid systems on an evolving surface, Quarterly of Applied Mathematics 75~(2)
  (2016) 359--389.
\newblock \href {https://doi.org/10.1090/qam/1452}
  {\path{doi:10.1090/qam/1452}}.

\bibitem{Reuther_MMS_2018}
S.~Reuther, A.~Voigt, Erratum: The interplay of curvature and vortices in flow
  on curved surfaces, Multiscale Modeling {\&} Simulation 16~(3) (2018)
  1448--1453.
\newblock \href {https://doi.org/10.1137/18m1176464}
  {\path{doi:10.1137/18m1176464}}.

\bibitem{Koba_2018}
H.~Koba, C.~Liu, Y.~Giga, Errata to {\textquotedblleft}energetic variational
  approaches for incompressible fluid systems on an evolving
  surface{\textquotedblright}, Quarterly of Applied Mathematics 76~(1) (2017)
  147--152.
\newblock \href {https://doi.org/10.1090/qam/1482}
  {\path{doi:10.1090/qam/1482}}.

\bibitem{Miura_2018}
T.-H. Miura, On singular limit equations for incompressible fluids in moving
  thin domains, Quarterly of Applied Mathematics 76~(2) (2017) 215--251.
\newblock \href {https://doi.org/10.1090/qam/1495}
  {\path{doi:10.1090/qam/1495}}.

\bibitem{Jankuhn_2018}
T.~Jankuhn, M.~A. Olshanskii, A.~Reusken, Incompressible fluid problems on
  embedded surfaces: Modeling and variational formulations, Interfaces and Free
  Boundaries 20~(3) (2018) 353--377.
\newblock \href {https://doi.org/10.4171/ifb/405} {\path{doi:10.4171/ifb/405}}.

\bibitem{Al-IzziAlexander_PRR_2023}
S.~C. Al-Izzi, G.~P. Alexander, Chiral active membranes: Odd mechanics,
  spontaneous flows, and shape instabilities, Physical Review Research 5~(4)
  (Dec. 2023).
\newblock \href {https://doi.org/10.1103/physrevresearch.5.043227}
  {\path{doi:10.1103/physrevresearch.5.043227}}.

\bibitem{Capovilla_2002}
R.~Capovilla, J.~Guven, Stresses in lipid membranes, Journal of Physics A:
  Mathematical and General 35~(30) (2002) 6233--6247.
\newblock \href {https://doi.org/10.1088/0305-4470/35/30/302}
  {\path{doi:10.1088/0305-4470/35/30/302}}.

\bibitem{bois2011pattern}
J.~S. Bois, F.~Jülicher, S.~W. Grill, Pattern formation in active fluids,
  Biophysical Journal 100~(3) (2011) 445a.
\newblock \href {https://doi.org/10.1016/j.bpj.2010.12.2620}
  {\path{doi:10.1016/j.bpj.2010.12.2620}}.

\bibitem{mietke2019minimal}
A.~Mietke, V.~Jemseena, K.~V. Kumar, I.~F. Sbalzarini, F.~Jülicher, Minimal
  model of cellular symmetry breaking, Physical Review Letters 123~(18) (2019)
  188101.
\newblock \href {https://doi.org/10.1103/physrevlett.123.188101}
  {\path{doi:10.1103/physrevlett.123.188101}}.

\bibitem{wittwer2023computational}
L.~D. Wittwer, S.~Aland, A computational model of self-organized shape dynamics
  of active surfaces in fluids, Journal of Computational Physics: X 17 (2023)
  100126.
\newblock \href {https://doi.org/10.1016/j.jcpx.2023.100126}
  {\path{doi:10.1016/j.jcpx.2023.100126}}.

\bibitem{Raetz_2006}
A.~Rätz, A.~Voigt, Pde’s on surfaces---a diffuse interface approach,
  Communications in Mathematical Sciences 4~(3) (2006) 575--590.
\newblock \href {https://doi.org/10.4310/cms.2006.v4.n3.a5}
  {\path{doi:10.4310/cms.2006.v4.n3.a5}}.

\bibitem{Nestler_2024}
M.~Nestler, A.~Voigt, A diffuse interface approach for vector-valued pdes on
  surfaces, arXiv (2023).
\newblock \href {https://doi.org/10.48550/ARXIV.2303.07135}
  {\path{doi:10.48550/ARXIV.2303.07135}}.

\bibitem{Giomi_2015}
L.~Giomi, Geometry and topology of turbulence in active nematics, Physical
  Review X 5~(3) (2015).
\newblock \href {https://doi.org/10.1103/physrevx.5.031003}
  {\path{doi:10.1103/physrevx.5.031003}}.

\bibitem{Pearce_2019}
D.~Pearce, P.~W. Ellis, A.~Fernandez-Nieves, L.~Giomi, Geometrical control of
  active turbulence in curved topographies, Physical Review Letters 122~(16)
  (Apr. 2019).
\newblock \href {https://doi.org/10.1103/physrevlett.122.168002}
  {\path{doi:10.1103/physrevlett.122.168002}}.

\bibitem{marchetti2013hydrodynamics}
M.~C. Marchetti, J.~F. Joanny, S.~Ramaswamy, T.~B. Liverpool, J.~Prost, M.~Rao,
  R.~A. Simha, Hydrodynamics of soft active matter, Reviews of Modern Physics
  85~(3) (2013) 1143--1189.
\newblock \href {https://doi.org/10.1103/revmodphys.85.1143}
  {\path{doi:10.1103/revmodphys.85.1143}}.

\bibitem{Doostmohammadi_2018}
A.~Doostmohammadi, J.~Ignés-Mullol, J.~M. Yeomans, F.~Sagués, Active
  nematics, Nature Communications 9~(1) (Aug. 2018).
\newblock \href {https://doi.org/10.1038/s41467-018-05666-8}
  {\path{doi:10.1038/s41467-018-05666-8}}.

\bibitem{Giomi_2014}
L.~Giomi, M.~J. Bowick, P.~Mishra, R.~Sknepnek, M.~Cristina~Marchetti, Defect
  dynamics in active nematics, Philosophical Transactions of the Royal Society
  A: Mathematical, Physical and Engineering Sciences 372~(2029) (2014)
  20130365.
\newblock \href {https://doi.org/10.1098/rsta.2013.0365}
  {\path{doi:10.1098/rsta.2013.0365}}.

\bibitem{Juelicher_2018}
F.~Jülicher, S.~W. Grill, G.~Salbreux, Hydrodynamic theory of active matter,
  Reports on Progress in Physics 81~(7) (2018) 076601.
\newblock \href {https://doi.org/10.1088/1361-6633/aab6bb}
  {\path{doi:10.1088/1361-6633/aab6bb}}.

\bibitem{Mirza2023}
W.~Mirza, A.~Torres-Sánchez, G.~Vilanova, M.~Arroyo, Variational formulation
  of active nematics: theory and simulation, ArXiv (2023).
\newblock \href {https://doi.org/10.48550/ARXIV.2306.01515}
  {\path{doi:10.48550/ARXIV.2306.01515}}.

\bibitem{Thampi_2016}
S.~Thampi, J.~Yeomans, Active turbulence in active nematics, The European
  Physical Journal Special Topics 225~(4) (2016) 651--662.
\newblock \href {https://doi.org/10.1140/epjst/e2015-50324-3}
  {\path{doi:10.1140/epjst/e2015-50324-3}}.

\bibitem{Rorai_2021}
C.~Rorai, F.~Toschi, I.~Pagonabarraga, Active nematic flows confined in a
  two-dimensional channel with hybrid alignment at the walls: A unified
  picture, Physical Review Fluids 6~(11) (Nov. 2021).
\newblock \href {https://doi.org/10.1103/physrevfluids.6.113302}
  {\path{doi:10.1103/physrevfluids.6.113302}}.

\bibitem{Leslie_1966}
F.~M. Leslie, Some constitutive equations for anisotropic fluids, The Quarterly
  Journal of Mechanics and Applied Mathematics 19~(3) (1966) 357--370.
\newblock \href {https://doi.org/10.1093/qjmam/19.3.357}
  {\path{doi:10.1093/qjmam/19.3.357}}.

\bibitem{Beris_1994}
A.~N. Beris, B.~J. Edwards, Thermodynamics of Flowing Systems: with Internal
  Microstructure, Oxford University Press, Oxford, 1994.
\newblock \href {https://doi.org/10.1093/oso/9780195076943.001.0001}
  {\path{doi:10.1093/oso/9780195076943.001.0001}}.

\bibitem{Parodi_1970}
O.~Parodi, Stress tensor for a nematic liquid crystal, Journal de Physique
  31~(7) (1970) 581--584.
\newblock \href {https://doi.org/10.1051/jphys:01970003107058100}
  {\path{doi:10.1051/jphys:01970003107058100}}.

\bibitem{shankar2018defect}
S.~Shankar, S.~Ramaswamy, M.~C. Marchetti, M.~J. Bowick, Defect unbinding in
  active nematics, Physical Review Letters 121~(10) (2018) 108002.
\newblock \href {https://doi.org/10.1103/physrevlett.121.108002}
  {\path{doi:10.1103/physrevlett.121.108002}}.

\bibitem{shankar2022topological}
S.~Shankar, A.~Souslov, M.~J. Bowick, M.~C. Marchetti, V.~Vitelli, Topological
  active matter, Nature Reviews Physics 4~(6) (2022) 380--398.
\newblock \href {https://doi.org/10.1038/s42254-022-00445-3}
  {\path{doi:10.1038/s42254-022-00445-3}}.

\bibitem{pearce2023passive}
D.~J.~G. Pearce, C.~Thibault, Q.~Chaboche, C.~Blanch-Mercader, Passive defect
  driven morphogenesis in nematic membranes, arXiv (2023).
\newblock \href {https://doi.org/10.48550/arXiv.2312.16654}
  {\path{doi:10.48550/arXiv.2312.16654}}.

\bibitem{nestler2019finite}
M.~Nestler, I.~Nitschke, A.~Voigt, A finite element approach for vector- and
  tensor-valued surface pdes, Journal of Computational Physics 389 (2019)
  48--61.
\newblock \href {https://doi.org/10.1016/j.jcp.2019.03.006}
  {\path{doi:10.1016/j.jcp.2019.03.006}}.

\bibitem{praetorius2022dune}
S.~Praetorius, F.~Stenger, Dune-curvedgrid - a dune module for surface
  parametrization, Archive of Numerical Software 6~(1) (2022).
\newblock \href {https://doi.org/10.11588/ans.2022.1.75917}
  {\path{doi:10.11588/ans.2022.1.75917}}.

\bibitem{munster2019attachment}
S.~Münster, A.~Jain, A.~Mietke, A.~Pavlopoulos, S.~W. Grill, P.~Tomancak,
  Attachment of the blastoderm to the vitelline envelope affects gastrulation
  of insects, Nature 568~(7752) (2019) 395--399.
\newblock \href {https://doi.org/10.1038/s41586-019-1044-3}
  {\path{doi:10.1038/s41586-019-1044-3}}.

\bibitem{Kruse_2005}
K.~Kruse, J.~F. Joanny, F.~Jülicher, J.~Prost, K.~Sekimoto, Generic theory of
  active polar gels: a paradigm for cytoskeletal dynamics, The European
  Physical Journal E 16~(1) (2005) 5--16.
\newblock \href {https://doi.org/10.1140/epje/e2005-00002-5}
  {\path{doi:10.1140/epje/e2005-00002-5}}.

\bibitem{Salbreux_2009}
G.~Salbreux, J.~Prost, J.~F. Joanny, Hydrodynamics of cellular cortical flows
  and the formation of contractile rings, Physical Review Letters 103~(5) (Jul.
  2009).
\newblock \href {https://doi.org/10.1103/physrevlett.103.058102}
  {\path{doi:10.1103/physrevlett.103.058102}}.

\bibitem{Thijssen_2020}
K.~Thijssen, L.~Metselaar, J.~M. Yeomans, A.~Doostmohammadi, Active nematics
  with anisotropic friction: the decisive role of the flow aligning parameter,
  Soft Matter 16~(8) (2020) 2065--2074.
\newblock \href {https://doi.org/10.1039/c9sm01963d}
  {\path{doi:10.1039/c9sm01963d}}.

\end{thebibliography}


\end{document}